\pdfoutput=1
\documentclass[jgrga]{AGUTeX}

\usepackage{lineno}

\usepackage{graphicx}
\authorrunninghead{HE ET AL.}

\titlerunninghead{CORONAL MAGNETIC FIELD OF AR 10930 FLARE}

\authoraddr{Corresponding author: Han He,
National Astronomical Observatories, Chinese Academy of Sciences,
20A Datun Rd., Chaoyang District, Beijing 100012, China.
(hehan@nao.cas.cn)}

\begin{document}

%
%

\title{Variations of the 3-D coronal magnetic field associated with the X3.4-class solar flare event of AR 10930}

%
%

\authors{Han He,\altaffilmark{1,2} Huaning Wang,\altaffilmark{1} Yihua Yan,\altaffilmark{1}
P. F. Chen\altaffilmark{2,3}, and Cheng Fang\altaffilmark{2,3}}

\altaffiltext{1}{Key Laboratory of Solar Activity, National Astronomical
Observatories, Chinese Academy of Sciences, Beijing, China.}

\altaffiltext{2}{Key Laboratory of Modern Astronomy and Astrophysics
(Nanjing University), Ministry of Education, Nanjing, China.}

\altaffiltext{3}{Department of Astronomy, Nanjing University, Nanjing,
China.}

%
%

\begin{abstract}
  The variations of the 3-D coronal magnetic fields associated with the X3.4-class flare of active region 10930 are studied in this paper. The coronal magnetic field data are reconstructed from the photospheric vector magnetograms obtained by the Hinode satellite and using the nonlinear force-free field extrapolation method developed in our previous work (He et al., 2011). The 3-D force-free factor $\alpha$, 3-D current density, and 3-D magnetic energy density are employed to analyze the coronal data. The distributions of $\alpha$ and current density reveal a prominent magnetic connectivity with strong negative $\alpha$ values and strong current density before the flare. This magnetic connectivity extends along the main polarity inversion line and is found to be totally broken after the flare. The distribution variation of magnetic energy density reveals the redistribution of magnetic energy before and after the flare. In the lower space of the modeling volume the increase of magnetic energy dominates, and in the higher space the decrease of energy dominates. The comparison with the flare onset imaging observation exhibits that the breaking site of the magnetic connectivity and site with the highest values of energy density increase coincide with the location of flare initial eruption. We conclude that a cramped positive $\alpha$ region appearing in the photosphere causes the breaking of the magnetic connectivity. A scenario for flare initial eruption is proposed in which the Lorentz force acting on the isolated electric current at the magnetic connectivity breaking site lifts the associated plasmas and causes the initial ejection.
\end{abstract}

%
%

\begin{article}

%
%

\section{Introduction}
The flare phenomenon in solar atmosphere is one of the most important research subjects of solar physics \citep{Benz_LRSP_2008, ShibataMagara_LRSP_2011, Hudson_SSRv_2011, FletcherEA_SSRv_2011}. It has close ties with coronal mass ejection (CME) \citep{ZhangLow_ARAA_2005, Chen_LRSP_2011, WebbHoward_LRSP_2012} as well as solar energetic particles \citep{Schwenn_LRSP_2006, Aschwanden_SSRv_2012}. These solar explosive phenomena are the main drivers of space weather \citep{Schwenn_LRSP_2006}.

The characteristics of solar flare activity have tight correlation with the 3-D magnetic field structure and its evolution in the solar atmosphere \citep{ShibataMagara_LRSP_2011}. It is believed that the major solar flare events are the results of the magnetic field reconnections happening in the solar atmosphere in solar active regions (ARs) \citep{PriestForbes_AARv_2002}. The 3-D coronal magnetic field distributions before and after flare eruption contain critical physical information to understand this change of magnetic connectivity.

Currently, although the coronal magnetic field can be diagnosed to some extent \citep[e.g.,][]{LinEA_ApJ_2004, Cargill_SSRv_2009, ChenEA_ApJ_2011}, it is still difficult to measure the 3-D coronal magnetic field distributions directly. We have to rely on physical models to reconstruct the 3-D coronal magnetic field data \citep{Sakurai_SolPhys_1981, Sakurai_SSRv_1989}.

The corona is composed of magnetized plasma with extremely low density and very high temperature compared to the photospheric values. In the inner corona, especially above active regions, the magnetic field dominates the plasma, other forces (such as gravitational force and plasma pressure) can be neglected. To achieve an equilibrium state, the electric current density vector
\begin{linenomath*}
\begin{equation}\label{equ:current-density}
  \textbf{j}=(1/4\pi)(\nabla\times \textbf{B})
\end{equation}
\end{linenomath*}
(in electromagnetic CGS units) must be everywhere parallel to the magnetic field vector $\textbf{B}$, then the Lorentz force vanishes. This physical situation can be described by the so-called nonlinear force-free field (NLFFF) model which consists of two main equations:
\begin{linenomath*}
\begin{equation}\label{equ:force-free-alpha}
    \nabla\times \textbf{B} = \alpha(\textbf{r})\textbf{B},
\end{equation}
\end{linenomath*}
\begin{linenomath*}
\begin{equation}\label{equ:alpha-constraint}
    \textbf{B}\cdot\nabla\alpha=0.
\end{equation}
\end{linenomath*}
The parameter $\alpha$ in equations (\ref{equ:force-free-alpha}) and (\ref{equ:alpha-constraint}) is called force-free factor. It has the dimension of $L^{-1}$ (reciprocal of length) and is a function of spatial location. As implied by equation (\ref{equ:alpha-constraint}), $\alpha$ is a constant along each magnetic field line.

The NLFFF model is a reasonable approximation to the quasi-equilibrium corona \citep{WiegelmannSakurai_LRSP_2012} and is commonly used for 3-D coronal magnetic field reconstruction above active regions \citep[e.g.,][]{SchrijverEA_ApJ_2008, HeWang_JGR_2008, JingEA_ApJ_2008, DeRosaEA_ApJ_2009, HeEA_JGR_2011, FuhrmannEA_AA_2011, LiuEA_ApSS_2012, GeorgoulisEA_ApJ_2012, JiangEA_ApJ_2012, InoueEA_ApJ_2012}. The 2-D photospheric vector magnetic field data (vector magnetograms), which can be measured sophisticatedly, act as the bottom boundary conditions in the NLFFF model and pose strong constraints to the upper coronal magnetic field distributions. Based on the observed photospheric vector magnetogram data and by solving equations (\ref{equ:force-free-alpha}) and (\ref{equ:alpha-constraint}) numerically (a process called NLFFF extrapolation \citep{SchrijverEA_SolPhys_2006, Regnier_MSAI_2007, Wiegelmann_JGRA_2008}), the reconstructed 3-D coronal magnetic field can be obtained, which is ready for the further analyses. By utilizing time series of photospheric vector magnetograms, the NLFFF model can also be used for tracing the evolution of the coronal magnetic field through a sequence of equilibria \citep{Sakurai_SSRv_1989}.

In our previous paper \citep{HeEA_JGR_2011}, we improved our NLFFF extrapolation method and code \citep{HeWang_JGR_2008} for the photospheric vector magnetograms obtained by the Spectro-Polarimeter (SP) focal plane instrument \citep{LitesEA_SolPhys_2013} of the Solar Optical Telescope (SOT) \citep{TsunetaEA_SolPhys_2008} aboard the Hinode satellite \citep{KosugiEA_SolPhys_2007} (Hinode/SOT-SP) and studied the time series evolution of the reconstructed 3-D coronal magnetic configurations of AR 10930 (NOAA ARs numbering system) when the active region was near the central meridian of the Sun's disk and at a relatively low level of activity. In this paper, we continue the work of \citet{HeEA_JGR_2011} and, by using the Hinode data and NLFFF model, study the variations of the 3-D coronal magnetic field associated with a major (X3.4-class) flare event of AR 10930.

The X3.4 flare of AR 10930 happened on 13 December 2006, which was captured and thoroughly observed by the instruments of Hinode \citep{WatanabeEA_SolPhys_2012} as well as other space-based and ground-based solar telescopes \citep{MinoshimaEA_ApJ_2009}. The evolution of AR 10930 and its relation to the X3.4 flare eruption have been analyzed by many authors from different viewpoints and approaches. For example, \citet{ZhangEA_ApJ_2007} examined the photospheric magnetic field and sunspot evolution of AR 10930 and related the X3.4 flare to the fast rotation of a positive polarity sunspot and the interaction between the ephemeral regions; \citet{MinChae_SolPhys_2009} further studied the rotating sunspot by using the higher resolution data of Hinode and related the sunspot rotation to the dynamic development of emerging twisted magnetic fields; \citet{ParkEA_ApJ_2010} investigated the time evolution of the coronal relative magnetic helicity in AR 10930 and related the occurrence of the X3.4 flare to the positive helicity injection into an existing system of negative helicity; \citet{InoueEA_ApJ_2012} analyzed the temporal evolution of the 3-D magnetic structure of AR 10930 through magnetic twist and related the X3.4 flare onset to a mix of differently twisted field lines and the strong currents contained in the central region near the polarity inversion line (PIL, the separation line between the positive and negative magnetic polarities of active region).

In this paper, we employ three quantitative physical measures, 3-D force-free factor $\alpha$, 3-D electric current density, and 3-D magnetic energy density, to analyze the internal variations of the coronal magnetic field associated with the X3.4 flare of AR 10930. Among the three coronal measures, the 3-D force-free factor $\alpha$ is introduced to reflect the magnetic connectivity in the coronal magnetic field. The electric current density distribution in AR 10930 has been discussed in the paper by \citet{SchrijverEA_ApJ_2008}. In this work, we extend the analysis to nine data sets of AR 10930 before and after the X3.4 flare to see time series variation of the current density distribution in the active region, which can be helpful to understanding the changes of internal fine structures of the coronal magnetic field. The magnetic energy stored in AR 10930 has been evaluated in several papers \citep[e.g.,][]{SchrijverEA_ApJ_2008, JingEA_ApJ_2010}. In this work, we emphasize the spatial distribution of the magnetic energy density and its time variation before and after the flare.

The contents of this paper are organized as follows: In section 2, we describe the data selection, projection effect correction, and alignment of the Hinode/SOT-SP photospheric vector magnetograms as well as the NLFFF extrapolation for reconstructing the 3-D coronal magnetic field associated with the X3.4 flare event of AR 10930. In section 3, we analyze the variations of the reconstructed 3-D coronal magnetic fields before and after the flare through the three quantitative physical measures, force-free factor $\alpha$, current density, and magnetic energy density. In section 4, we compare the distribution variations of the coronal physical measures before and after the flare with the imaging observation by Hinode/SOT during the flare onset phase to investigate the relation between the internal variations of the coronal magnetic field and the flare initial eruption. In section 5, we make further discussions on the results of sections 3 and 4. In section 6, we give summary and conclusion.

\section{Reconstruction of the 3-D Coronal Magnetic Field}

\subsection{Photospheric Vector Magnetograms}
According to the Solar Event Reports issued by the Space Weather Prediction Center of NOAA (NOAA/SWPC) and the solar soft X-ray flux data obtained by Geostationary Operational Environmental Satellites 11 and 12 (GOES 11 and GOES 12) in the 1.0--8.0~{\AA} wavelength band, the beginning time and the maximum time of the X3.4-class X-ray flare of AR 10930 are 13 December 2006 02:14 UT and 02:40 UT, respectively. Before this flare, the active region was at a relatively low level of activity for several days (8--12 December 2006) with no major flare events. We study the variations of the coronal magnetic field of AR 10930 from 11 December 2006, when the active region was just in the central area of the Sun's disk, until the end of the X3.4 flare event on 13 December 2006.

Nine photospheric vector magnetograms of AR 10930 obtained by the Hinode/SOT-SP instrument within the 3~days between 11 and 13 December 2006 are selected for the coronal magnetic field reconstruction, six magnetograms for the preflare phase and three magnetgrams for the postflare phase. All nine magnetograms were observed in the Fast Map operation mode of Hinode/SOT-SP \citep{TsunetaEA_SolPhys_2008, LitesEA_SolPhys_2013}, which takes 63 min to complete a scan of $295'' \times 162''$ field of view (FOV). The Fast Map mode of Hinode/SOT-SP is preferred in this work because it possesses the best balance between larger FOV and higher cadence for the extrapolation and time series evolution studies of AR 10930 \citep{HeEA_JGR_2011}.

Table 1 gives the basic information (observation time, FOV, coordinates, preflare/postflare) of the nine selected Hinode/SOT-SP photospheric vector magnetograms. For ease of identification, the nine magnetograms are labeled from M1 to M9, respectively, as shown in Table 1. Figure 1 illustrates the observation times of the nine selected Hinode/SOT-SP magnetograms on the graph of the solar soft X-ray flux curves recorded by GOES 11 and GOES 12 satellites in the 1.0--8.0~{\AA} wavelength band. The graph of the solar X-ray flux covers the time period from 10 December 2006 00:00 UT to 15 December 2006 00:00 UT and includes the X3.4 flare event on 13 December 2006. In Figure 1, the observation times of the nine selected magnetograms are indicated by nine vertical bars which are marked from M1 to M9, respectively, according to the labels of the magnetograms. The width of each vertical bar represents the duration of observation period (about 63 min) for each magnetogram. As demonstrated in Figure 1, M1--M6 are preflare magnetograms, M7--M9 are postflare magnetograms.

\begin{table*}
  \caption{Basic Information of the Nine Hinode/SOT-SP Photospheric Vector Magnetograms Selected for Reconstructing the 3-D Coronal Magnetic Field Associated With the X3.4 Flare Event of AR 10930}
  \centering
  \begin{tabular}{cccccccc}
  \tableline
  &  &  Original FOV\tablenotemark{c} & XCEN\tablenotemark{d} & YCEN\tablenotemark{e} & LONCEN\tablenotemark{f} & LATCEN\tablenotemark{g} \\
  Label\tablenotemark{a} & Time of Observation\tablenotemark{b} & (arcsec) & (arcsec) & (arcsec) & (deg) & (deg) & Preflare/Postflare\tablenotemark{h} \\
  \tableline
  M1 & 2006-12-11 08:00:04 UT & $295 \times 162$ &  $-$20.07 & $-$88.17 & $-$1.187 & $-$5.609 & Preflare \\
  M2 & 2006-12-11 23:10:05 UT & $295 \times 162$ &  122.01 & $-$87.22 &  7.232 & $-$5.630 & Preflare \\
  M3 & 2006-12-12 03:50:05 UT & $295 \times 162$ &  165.15 & $-$85.80 &  9.809 & $-$5.567 & Preflare \\
  M4\tablenotemark{i} & 2006-12-12 15:30:08 UT & $295 \times 162$ &  272.81 & $-$85.80 & 16.346 & $-$5.613 & Preflare \\
  M5 & 2006-12-12 17:40:05 UT & $295 \times 162$ &  292.69 & $-$86.33 & 17.575 & $-$5.652 & Preflare \\
  M6 & 2006-12-12 20:30:05 UT & $295 \times 162$ &  317.88 & $-$85.71 & 19.143 & $-$5.624 & Preflare \\
  M7 & 2006-12-13 04:30:05 UT & $295 \times 162$ &  386.34 & $-$82.85 & 23.481 & $-$5.478 & Postflare \\
  M8 & 2006-12-13 07:50:05 UT & $295 \times 162$ &  413.46 & $-$83.15 & 25.240 & $-$5.503 & Postflare \\
  M9 & 2006-12-13 16:21:04 UT & $295 \times 162$ &  486.85 & $-$83.92 & 30.142 & $-$5.561 & Postflare \\
  \tableline
  \end{tabular}
  \tablenotetext{a}{The nine Hinode/SOT-SP magnetograms of AR 10930 are labeled from M1 to M9, respectively, for ease of identification.}
  \tablenotetext{b}{All nine magnetograms were observed in the Fast Map operation model of Hinode/SOT-SP \citep{TsunetaEA_SolPhys_2008, LitesEA_SolPhys_2013} which takes 63 min to complete a scan of the whole $295'' \times 162''$ FOV. The times listed in this column are the start time of observation. Dates are formatted as year/month/day.}
  \tablenotetext{c}{The original magnetograms need to be corrected for the projection effects before doing coronal magnetic field reconstruction. The FOV of the corrected magnetograms is $290'' \times 160''$, with the distortion edges being cut out to acquire aligned and uniform maps. }
  \tablenotetext{d}{XCEN is the $X$ coordinate of the magnetogram's center point in the heliocentric coordinate system. The coordinate data of the Hinode/SOT-SP magnetograms listed here (XCEN, YCEN, LONCEN, and LATCEN) are adjusted by comparing with the SOHO/MDI full-disk magnetograms \citep{ScherrerEA_SolPhys_1995} through cross-correlation technique \citep{FisherWelsch_ASPC_2008}, which are sightly different from the coordinate values contained in the Level 1 and Level 2 data of Hinode/SOT-SP \citep{CentenoEA_ASPC_2009, LitesIchimoto_SolPhys_2013}.}
  \tablenotetext{e}{YCEN is the $Y$ coordinate of the magnetogram's center point in the heliocentric coordinate system.}
  \tablenotetext{f}{LONCEN is the longitude value of the magnetogram's center point in the heliographic coordinate system, positive to the west, negative to the east.}
  \tablenotetext{g}{LATCEN is the latitude value of the magnetogram's center point in the heliographic coordinate system, positive to the north, negative to the south.}
  \tablenotetext{h}{The X3.4 flare event happened on 13 December 2006 with the beginning time and the maximum time being 02:14 UT and 02:40 UT, respectively. Then M1--M6 are preflare magnetograms, M7--M9 are postflare magnetograms.}
  \tablenotetext{i}{Certain columns of pixels were missing in the original Hinode/SOT-SP data of M4, which are patched up via interpolation. The missing pixels are located at the left (east) side of the magnetogram and in the weak field region, which has little impact on the result of coronal magnetic field reconstruction.}
\end{table*}

Figure 2 displays the locations and coverage areas of the nine selected Hinode/SOT-SP magnetograms of AR 10930 on the disk of the Sun. In Figure 2, the center points of the nine magnetograms (representing the locations of the magnetograms) are indicated by nine diamond symbols lined up from the central region of the Sun's disk to the west (M1 at the easternmost point, M9 at the westernmost point), the coverage areas of the magnetograms are represented by white rectangles (only shown for M1 and M9 to keep the diagram concise, the other magnetograms having the same FOV). The background full-disk line-of-sight (LOS) magnetogram employed in Figure 2 was observed by the Michelson Doppler Imager (MDI) \citep{ScherrerEA_SolPhys_1995} aboard the Solar and Heliospheric Observatory (SOHO) spacecraft \citep{DomingoEA_SolPhys_1995} (SOHO/MDI) at 13 December 2006 01:36:27 UT (38 min before the start of the X3.4 flare of AR 10930).

The Level 0 data \citep{MatsuzakiEA_SolPhys_2007} associated with the nine selected magnetograms are the original Stokes spectral data of Fe \textsc{i} lines at 6301.5~{\AA} and 6302.5~{\AA} \citep{TsunetaEA_SolPhys_2008, LitesEA_SolPhys_2013} acquired by Hinode/SOT-SP; the Level 1 data of Hinode/SOT-SP are the calibrated Stokes profile data \citep{LitesIchimoto_SolPhys_2013}; the Level 2 data are the vector magnetic field data after the inversion of Stokes profiles \citep{LitesEA_MmSAI_2007}. Our studies in this paper are mainly based on the Level 2 data of Hinode/SOT-SP. The Level 0 and Level 1 data associated with the nine selected magnetograms are also employed to obtain the detailed coordinates information of the magnetograms and to help determine the locations of the missing pixels (see footnote i of Table 1) in the Level 2 data.

The $180^\circ$ ambiguities \citep{MetcalfEA_SolPhys_2006, LekaEA_SolPhys_2009} in the transverse components of the nine Hinode/SOT-SP photospheric vector magnetograms were resolved by using the NPFC (nonpotential field calculation) disambiguation code developed by \citet{Georgoulis_ApJ_2005}.

\subsection{Projection Effect Correction and Alignment of the Photospheric Vector Magnetograms}
Considering that the X3.4 flare of AR 10930 occurred at the west side of the Sun's disk (about W$22.5^{\circ}$ in the heliographic coordinate system, see Figure 2), the projection effects in the Hinode/SOT-SP vector magnetograms need to be corrected both for the shapes of the magnetograms and for the directions of the magnetic field vectors before doing coronal magnetic field reconstruction (NLFFF extrapolation).

When observing an active region, the pointing of the Hinode satellite is set up to track the active region continuously with the speed of the solar differential rotation \citep{TsunetaEA_SolPhys_2008}. We conduct the projection effect corrections for the Hinode/SOT-SP magnetograms accordingly. That is, the corrected vector magnetograms are what we see from just above the tracking point of the active region. By matching the coordinates of the tracking point in the process of projection effect correction, the alignment of the corrected magnetograms can also be achieved.

In the practical operation of the projection effect correction, to fully utilize the original resolution of the Hinode/SOT-SP Fast Map data ($0.295''\times0.317''$/pixel \citep{KuboEA_PASJ_2007}), we first adopt $0.25''$/pixel resampling resolution to produce the projection-effect-corrected vector magnetograms. The FOV of the corrected magnetograms is $290'' \times 160''$ ($1160 \times 640$ pixels) with the distortion edges being cut out to acquire an aligned and uniform time series. Then the $0.25''$/pixel data of the corrected magnetograms are rebinned to $1''$/pixel data ($290 \times 160$ grid and $290'' \times 160''$ FOV) which act as the actual input data for the NLFFF extrapolations (see section 2.3).

\begin{figure*}
  \centering
  \noindent\includegraphics[width=30pc]{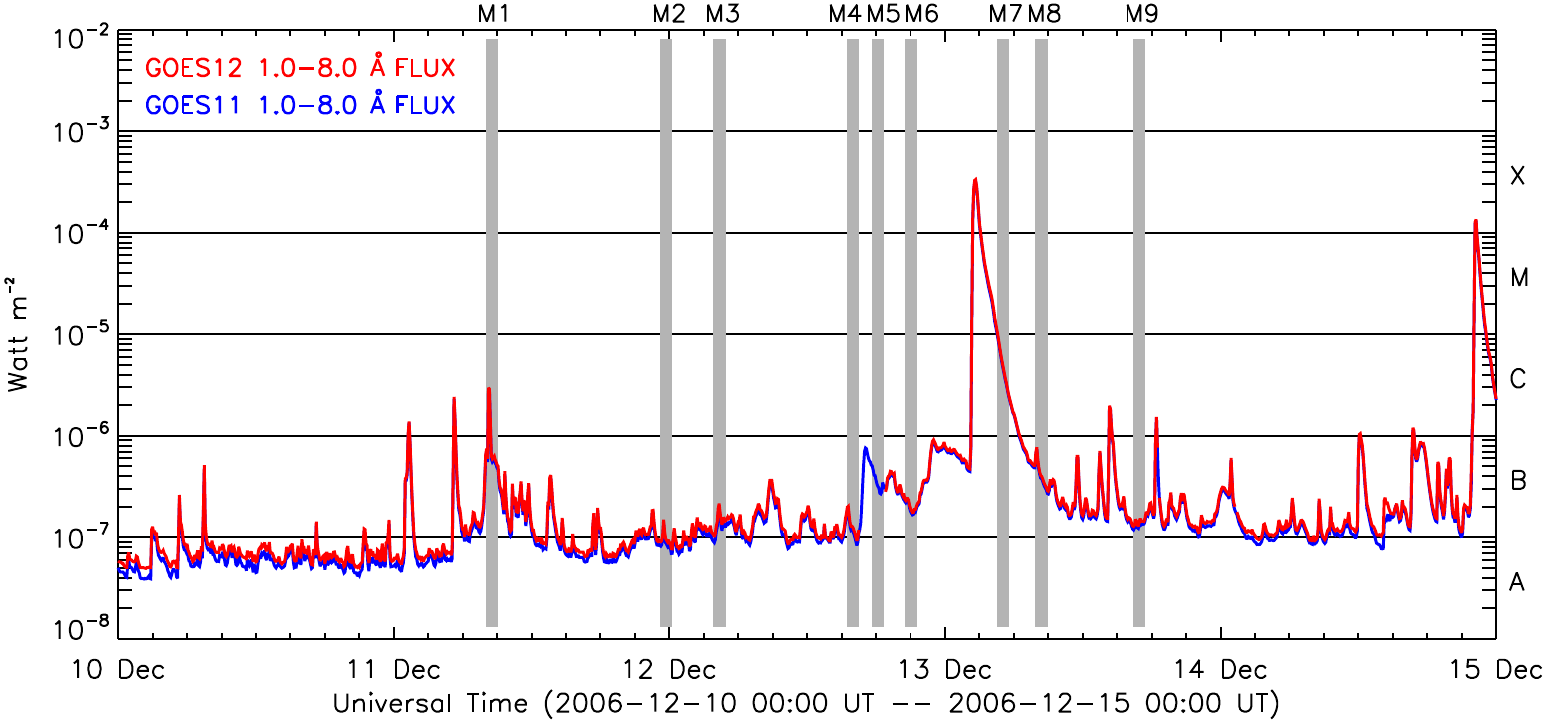}
  \caption{Diagram illustrating the observation times of the nine Hinode/SOT-SP photospheric vector magnetograms selected for reconstructing the 3-D coronal magnetic field associated with the X3.4 flare event of AR 10930. The graph of the solar soft X-ray flux curves from 10 December 2006 00:00 UT to 15 December 2006 00:00 UT is included for reference, which covers the X3.4 flare event on 13 December 2006. The observation times of the nine selected magnetograms are indicated by nine vertical bars, which are marked from M1 to M9, respectively, according to the labels of the magnetograms (see Table~1). The width of each vertical bar represents the duration of observation period (about 63 min) for each magnetogram. The solar soft X-ray flux data were recorded by GOES 11 and GOES 12 satellites in the 1.0--8.0~{\AA} wavelength band.}
\end{figure*}

\begin{figure*}
  \centering
  \noindent\includegraphics[width=20pc]{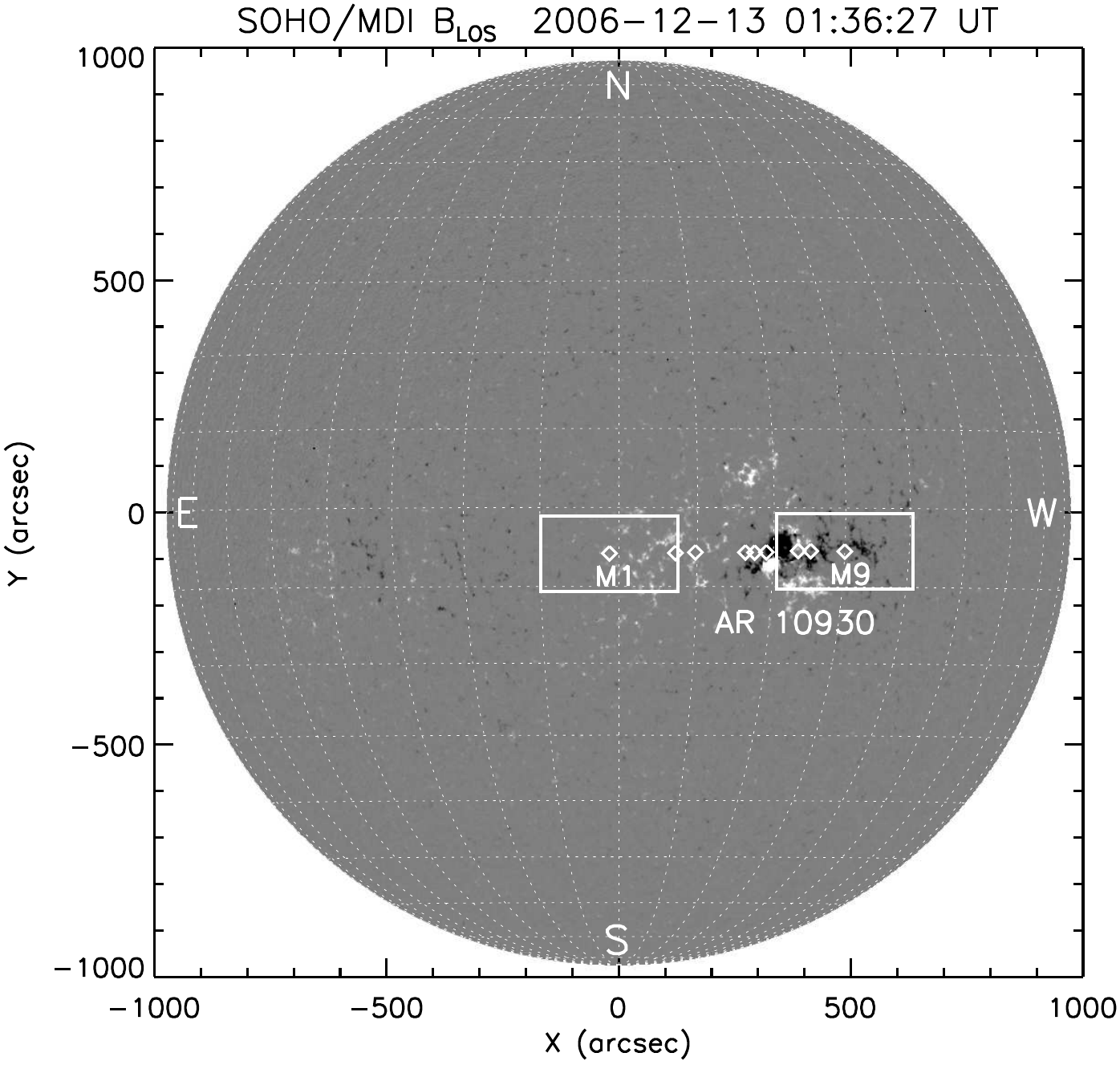}
  \caption{Diagram illustrating the locations and coverage areas of the nine Hinode/SOT-SP photospheric vector magnetograms selected for reconstructing the 3-D coronal magnetic field associated with the X3.4 flare event of AR 10930. The locations (center points) of the nine magnetograms are indicted by nine diamond symbols lined up from the central region of the Sun's disk to the west (see Table 1 for exact values of coordinates). M1 is the earliest magnetogram which is located at the easternmost point of the nine magnetograms, M9 is the last magnetogram which is locate at the westernmost point. The coverage areas of M1 and M9 are represented by two white rectangles, the other magnetograms have the same FOV ($295'' \times 162''$). The background full-disk line of sight (LOS) magnetogram was observed by SOHO/MDI at 13 December 2006 01:36:27 UT (38 min before the start of the X3.4 flare of AR 10930), white represents positive polarity, and black represents negative polarity.}
\end{figure*}

\subsection{NLFFF Extrapolation}
The NLFFF extrapolation method employed to reconstruct the 3-D coronal magnetic field from the nine selected Hinode/SOT-SP photospheric vector magnetograms is the same as the method described and applied in our previous paper \citep{HeEA_JGR_2011}, which is a NLFFF extrapolation scheme based on the direct boundary integral equation (DBIE) formulation \citep{YanLi_ApJ_2006} (see also the papers by \citet{HeWang_MNRAS_2006, HeWang_JGR_2008} for more details about the development path of our method).

In this work, we adopt $1''$/pixel resolution (uniform in $X$, $Y$, and $Z$ direction, $1''\sim 714$km) to perform the NLFFF extrapolation calculations (a relatively higher resolution than that in our previous work \citep{HeEA_JGR_2011}). The input 2-D photospheric vector magnetograms are $290 \times 160$ grid (rebinned data, see section 2.2) with $290'' \times 160''$ FOV (in $X$ and $Y$ direction, respectively). The output 3-D coronal magnetic fields from the NLFFF extrapolation code are $290 \times 160 \times 160$ grid (in $X$, $Y$, and $Z$ direction, respectively) with the photospheric magnetogram being retained in the output 3-D data as the bottom layer (layer 0) of the data cube. Our following analyses are mainly based on the nine sets of 3-D coronal magnetic configurations extrapolated from the nine projection-effect-corrected and aligned Hinode/SOT-SP photospheric vector magnetograms of AR 10930. (The force-freeness of the reconstructed coronal magnetic fields by using our NLFFF extrapolation code for the Hinode/SOT-SP photospheric vector magnetograms has been carefully checked and verified in our previous paper \citep{HeEA_JGR_2011}.)

\section{Analyses of the 3-D Coronal Magnetic Field}

\subsection{Variation of the Photospheric Magnetic Field}

The photospheric magnetic field of AR 10930 evolved continuously in the three-day time period (11--13 December 2006) when the nine selected Hinode/SOT-SP vector magnetograms were observed before and after the X3.4 flare. Figure 3 displays the $B_z$ components of the nine projection-effect-corrected and aligned Hinode/SOT-SP magnetograms in full resampling resolution ($0.25''$/pixel, see section 2.2) and full FOV ($290'' \times 160''$), and Figure 4 shows the evolution of the vector magnetic field in the central area (strong field region) of the active region. (Hereafter, we use the same $X$ and $Y$ coordinates in the figures, in which the coordinate values are counted from the lower left corner of the full FOV of the magnetograms.)

As shown in Figures 3 and 4, the location of the main negative polarity of AR 10930 in the north was almost unchanged during the 3 day evolution, while the main positive polarity in the south rotated counterclockwise and moved eastward all the time, which reflects the behavior of the dense plasma flow beneath the photosphere \citep{Fan_LRSP_2009}. The transverse component $B_t$ of the vector magnetograms also changed continuously with the development of magnetic shear \citep{ZhangEA_ApJ_2007} along with the motion of the main positive polarity (see Figure 4).

\begin{figure*}
  \centering
  \noindent\includegraphics[width=41pc]{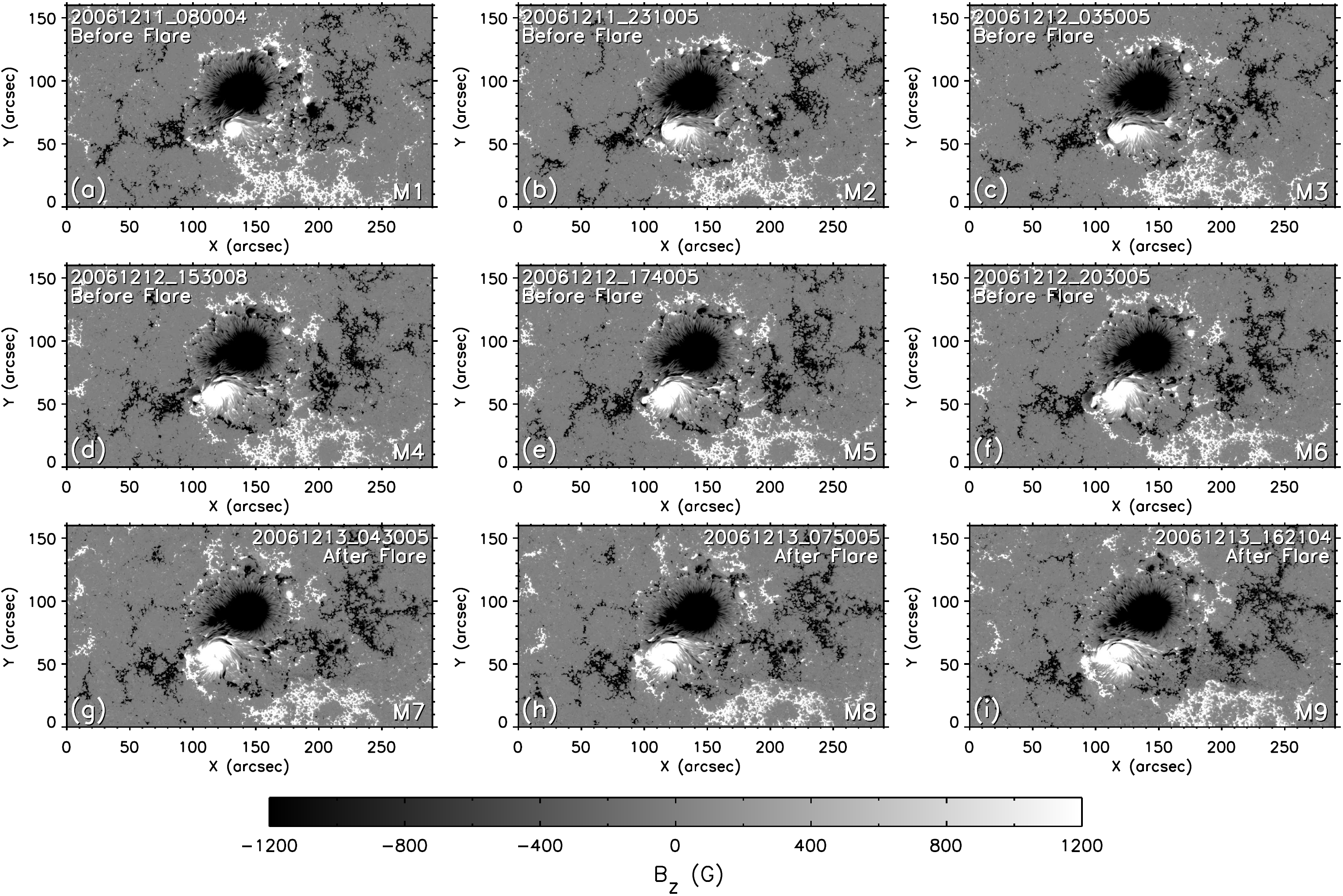}
  \caption{$B_z$ component images of the nine projection-effect-corrected and aligned Hinode/SOT-SP magnetograms of AR 10930 observed before and after the X3.4 flare. The images are displayed in full resampling resolution ($0.25''$/pixel) and full FOV ($290'' \times 160''$). The gray scale for $B_z$ saturates at $\pm 1200$~G, white represents positive polarity, and black represents negative polarity. Hereafter, we use the same $X$ and $Y$ coordinates in the figures, in which the coordinate values are counted from the lower left corner of the full FOV of the magnetograms. (Please enlarge electronic version of this figure to see detail.)}
\end{figure*}

The rotation of the main positive polarity in AR 10930 has been quantitatively studied by \citet{ZhangEA_ApJ_2007} and \citet{MinChae_SolPhys_2009} through the rotation angle of the corresponding sunspot. In the work of \citet{ZhangEA_ApJ_2007}, they examined the counterclockwise rotation of the positive polarity sunspot in AR 10930 based on the white light observations by the Transition Region and Coronal Explorer (TRACE) satellite \citep{HandyEA_SolPhys_1999} and found a total rotation angle of $240^\circ$ from the end of 10 December 2006 to 13 December 2006. \citet{MinChae_SolPhys_2009} employed a new technique and the higher-resolution data (G band images taken by Hinode/SOT \citep{TsunetaEA_SolPhys_2008}) to study the behavior of the rotating positive polarity sunspot and reported a total counterclockwise rotation angle of $540^\circ$ during 9--13 December 2006.

In this work, we focus on the variations of the magnetic field before and after the X3.4 flare of AR 10930. The $B_z$ component images of M6 and M7 (observed just before and after the flare eruption, see Figure 1) in the central area of the active region are compared in Figure 5 to show the changes of the photospheric magnetic fields before and after the flare directly. Figures 5a and 5b are gray scale images of $B_z$ using the full resolution data ($0.25''$/pixel, see section 2.2) of M6 and M7. Figures 5c and 5d are contour plots of $B_z$ using the rebinned data ($1''$/pixel, see section 2.2) of M6 and M7, which can display the coverage areas of the positive polarity (white contours) and the negative polarity (black contours) in the magnetograms more clearly. The counterclockwise rotation of the main positive polarity in AR 10930 is represented by the curved thick black arrows in Figures 5a and 5c.

Along with the photospheric motion in AR 10930, an apparent variation in the photospheric magnetic field between M6 and M7 is the appearance of a narrow patch with negative flux and kG magnitude of $B_z$ in the positive polarity region of M7, which is located right below the main PIL (polarity inversion line between the main positive and negative polarities) of M7 and is indicated by the thick white arrows in Figures 5b and 5d. We will further discuss this negative flux patch in the following sections.

The variation of the photospheric magnetic field can also be shown by the estimation of the total unsigned magnetic flux through the photosphere, which will be given in section 3.4.1 along with the evaluation of the total magnetic energy.

\begin{figure*}
  \centering
  \noindent\includegraphics[width=42pc]{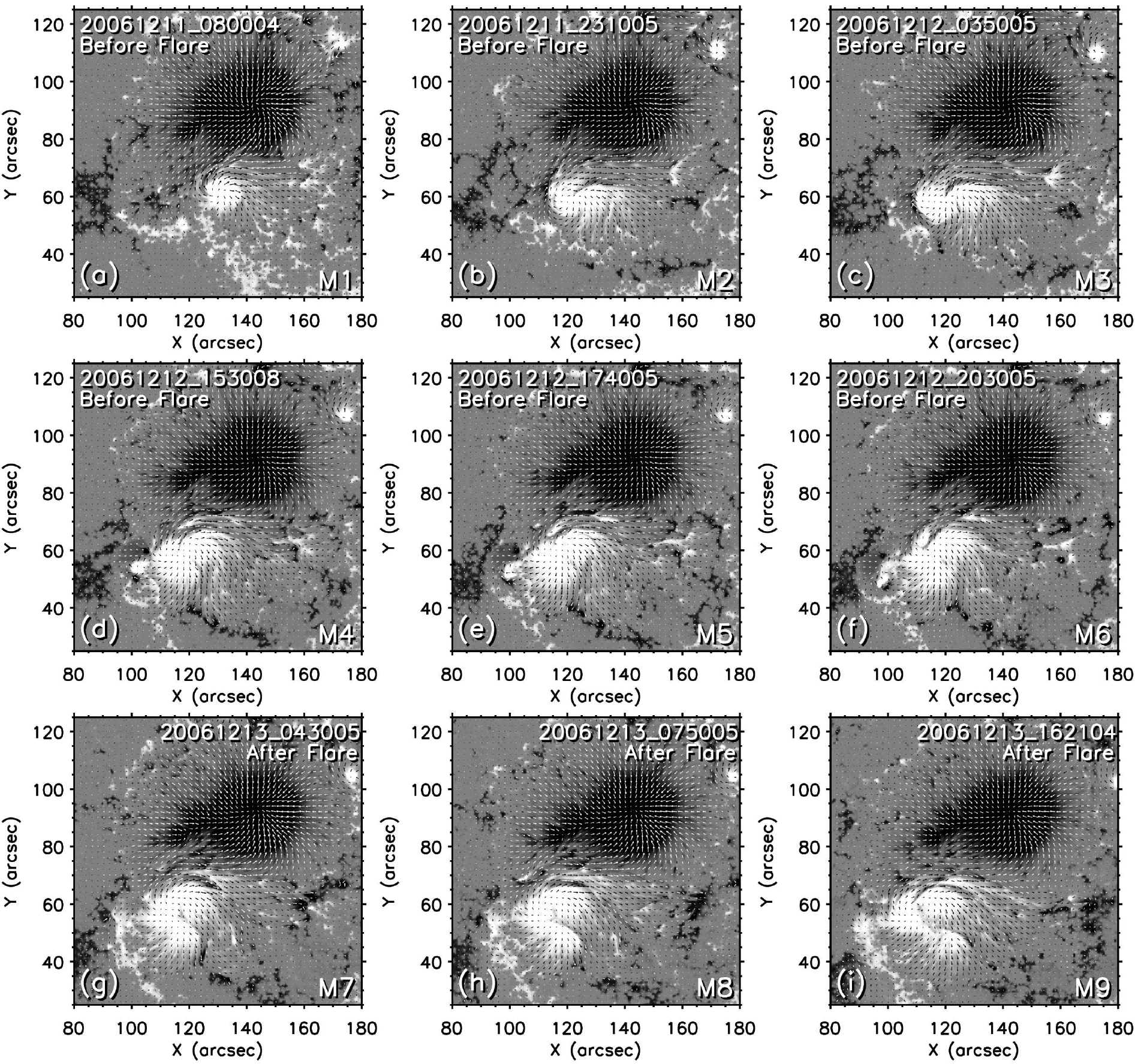}
  \caption{Images of the vector magnetic fields in the central areas of the nine projection-effect-corrected and aligned Hinode/SOT-SP magnetograms of AR 10930 observed before and after the X3.4 flare. The $B_z$ components of the magnetograms are displayed via gray scale images in full resampling resolution ($0.25''$/pixel), white represents positive polarity, black represents negative polarity, and the gray scale for $B_z$ saturates at $\pm 1800$ G. Small arrows overlying the $B_z$ images represent the transverse component $B_t$ of the magnetograms. The length of a cell in the arrow grids corresponds to 2250~G of $B_t$. (Please enlarge electronic version of this figure to see detail.)}
\end{figure*}

\subsection{Variation of the Field Line Configuration}
The field line configurations of the nine sets of reconstructed 3-D coronal magnetic field overlying the corresponding $B_z$ component images of photospheric magnetograms are shown in Figure 6 in full FOV ($290'' \times 160''$). The field lines displayed in Figure 6 are traced and plotted in the volume between layer 0 (bottom layer with photospheric magnetogram data) and layer 29 of the 3-D coronal data grid ($1''$/pixel, see section 2.3), the starting points of field line tracing are evenly spaced on layer 0. Open field lines (field lines that leave the given volume) in Figure 6 are plotted in black color, and closed field lines (field lines with both foot points being anchored at the layer 0) are plotted in white color.

The nine field line images in Figure 6 are arranged from left to right and from top to bottom in the order of their associated magnetograms to give a full view of the time series variation of the coronal magnetic field configurations as the responses to the magnetic field changes in the photosphere \citep{HeEA_JGR_2011}. In Figures 6f and 6g, it can be seen that the field line configurations associated with M6 and M7, which were observed just before and after the X3.4 flare of AR 10930, show distinct difference in the central area of the active region. In the following subsections, we use quantitative physical measures to analyze the variations of the coronal magnetic field in the core of the active region before and after the flare.

\subsection{Variations of the Force-Free Factor $\alpha$ and Electric Current Density Distribution}

The quantitative physical measure force-free factor $\alpha$ is defined in equation (\ref{equ:force-free-alpha}), which is a scalar function of spatial location and represents the nonpotential nature of active regions when different from zero. An important characteristic of the force-free factor $\alpha$ is that the value of $\alpha$ is a constant along each field line (in the context of NLFFF model) as implied by equation (\ref{equ:alpha-constraint}). Then the 3-D distribution of $\alpha$ can reflect the magnetic connectivity in the 3-D coronal magnetic field.

\begin{figure*}
  \centering
  \noindent\includegraphics[width=30pc]{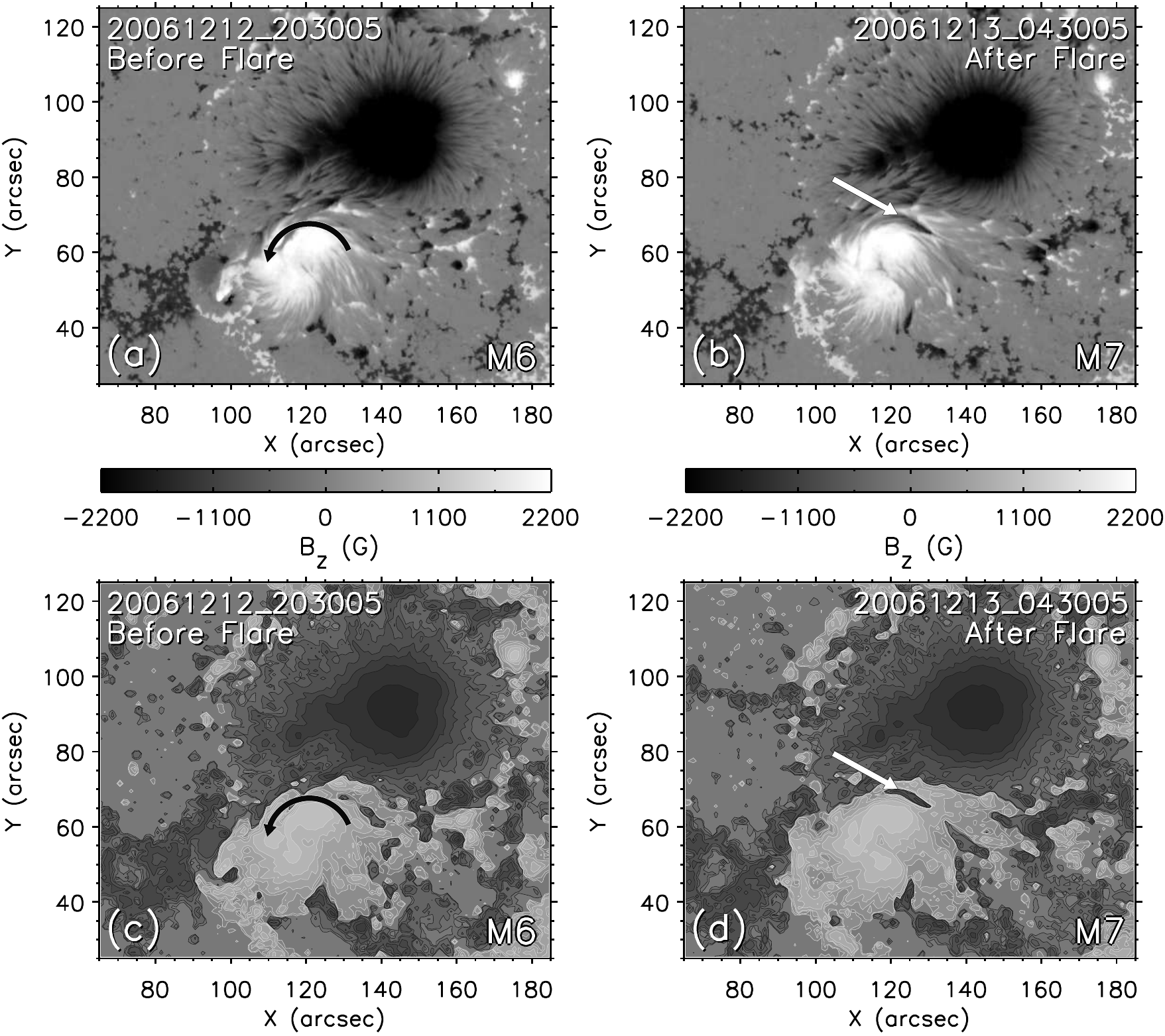}
  \caption{Comparison between the $B_z$ component images of the Hinode/SOT-SP magnetograms M6 and M7 in the central area of AR 10930. M6 were observed just before the X3.4 flare of AR 10930, and M7 were observed just after the flare. (a and b) Gray scale images of $B_z$ using the full resolution data ($0.25''$/pixel) of M6 and M7. White represents positive polarity, black represents negative polarity, and the gray scale for $B_z$ saturates at $\pm 2200$ G. (c and d) Contour images of $B_z$ using the rebinned data ($1''$/pixel) of M6 and M7, which can display the coverage areas of positive polarity (white contours) and negative polarity (black contours) in the magnetograms more clearly. The contour levels for $B_z$ are $\pm$50, 100, 300, 500, 1000, 1500, 2000, and 3000~G. The curved thick black arrows in Figures 5a and 5c represent the counterclockwise rotation of the main positive polarity in the active region. The thick white arrows in Figures 5b and 5d indicate a narrow patch that appears in M7 with negative flux and kG magnitude of $B_z$.}
\end{figure*}

The quantitative physical measure electric current density $\textbf{j}$ is expressed in equation (\ref{equ:current-density}), which is an important quantity for describing the magnetic activity in active regions \citep{LekaBarnes_ApJ_2007}. In the context of NLFFF model, the orientation of the current density vector $\textbf{j}$ is parallel to the magnetic field vector $\textbf{B}$, then the distribution of current density can also be helpful to understanding the internal fine structures of the coronal magnetic field \citep{SchrijverEA_ApJ_2008}.

In this subsection, we use the distribution of force-free factor $\alpha$ together with the distribution of electric current density to analyze the variation of the magnetic connectivity in the coronal magnetic fields of AR 10930 before and after the X3.4 flare.

\subsubsection{Force-Free Factor $\alpha$ Distribution}
The 3-D spatial distribution data of the force-free factor $\alpha$ were calculated from the nine sets of reconstructed 3-D coronal magnetic field data ($1''$/pixel, see section 2.3) by using the following equation \citep{HeEA_JGR_2011}:
\begin{linenomath*}
\begin{equation}\label{eq:cal-3D-alpha}
  \alpha= \frac{(\nabla\times \textbf{B})\cdot \textbf{B}}{B^2},
\end{equation}
\end{linenomath*}
which can be deduced from equation (\ref{equ:force-free-alpha}).

By examining the calculated 3-D data of $\alpha$, it is found that the typical absolute values of $\alpha$ in the coronal magnetic field of AR 10930 are about $0.1$ Mm$^{-1}$, which is consistent with the distribution range of $\alpha$ values in the literatures \citep[e.g.,][]{PevtsovEA_ApJ_1995, Leka_SolPhys_1999}. The $\alpha$ values with this order of magnitude mainly spread in the regions around the PILs and not far from the photosphere (corresponding to the low-lying sheared magnetic field lines over the PILs). In the central area of the active region, the distribution of $\alpha$ values presents sharp decline with increase in height (see Appendix~A for illustrations of $\alpha$ vertical distributions in AR 10930).

\begin{figure*}
  \centering
  \noindent\includegraphics[width=41pc]{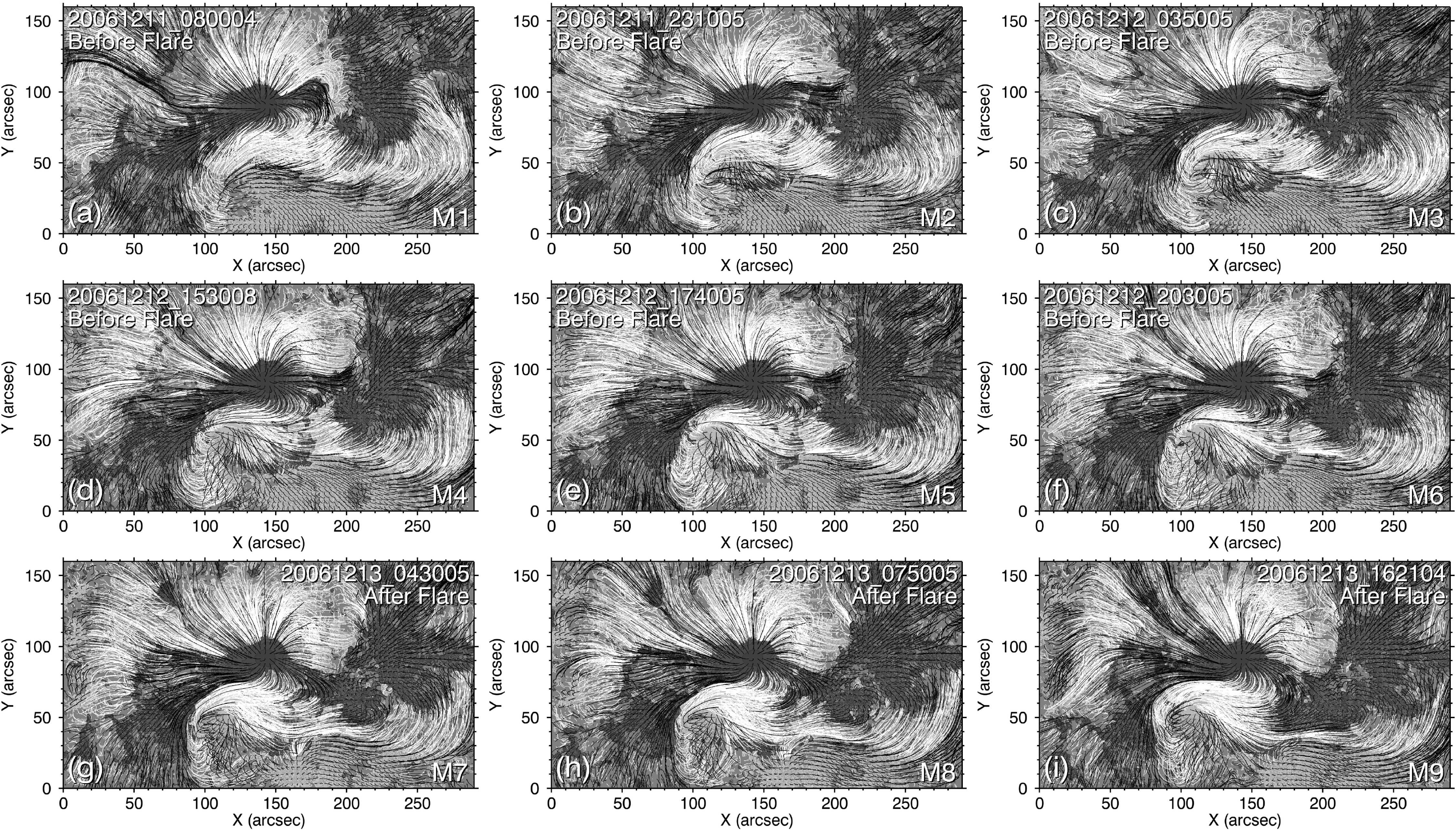}
  \caption{Field line images of the nine sets of reconstructed 3-D coronal magnetic field associated with the X3.4 flare event of AR 10930. The field lines are displayed overlying the $B_z$ contour images of the corresponding photospheric magnetograms in full FOV ($290'' \times 160''$). The field lines are traced in the volume between layer 0 and layer 29 ($290 \times 160 \times 30$ pixel volume) of the 3-D coronal data grid ($1''$/pixel), and the starting points of field line tracing are evenly spaced on the bottom layer (layer 0). Open field lines (field lines that leave the given volume) are plotted in black, and closed field lines (field lines with both foot points being anchored at the layer 0) are in white. (Please enlarge electronic version of this figure to see detail.)}
\end{figure*}

We use the data of $\alpha$ on layer 2 of the nine reconstructed 3-D coronal magnetic field data sets to demonstrate the time variation of the $\alpha$ distributions before and after the X3.4 flare of AR 10930. The resulting images (horizontal distribution of $\alpha$ on layer 2) over the full FOV ($290'' \times 160''$) are displayed in Figure 7. (More $\alpha$ distribution images on layers of different heights can be found in Appendix~A.)

From Figure 7 it can be seen that, in the central area of AR 10930, the local distributions of $\alpha$ are dominated by negative values (represented by red color in Figure 7). The negative values of $\alpha$ mean that the current density vector and the coronal magnetic field vector have opposite directions (in the context of NLFFF model). In the marginal area of the active region with weak magnetic field ($|\textbf{B}|$ approaching zero), the values of $\alpha$ are noisy as shown in Figure 7 and may not be reliable.

\subsubsection{Electric Current Density Distribution}
The 3-D data of the electric current density were calculated from the nine sets of reconstructed 3-D coronal magnetic field data ($1''$/pixel, see section 2.3) by using equation (\ref{equ:current-density}). We introduce a vertically averaged intensity map, $I(x, y)$, as in the paper by \citet{HeEA_JGR_2011} to display the current density magnitude distribution in the coronal magnetic field, which is defined as
\begin{linenomath*}
\begin{equation}
  I(x, y)=\frac{1}{L_z}\int_{0}^{L_z}\left| \textbf{j} (x, y, z)\right| \textrm{d}z,
\label{equ:averaged-current-density}
\end{equation}
\end{linenomath*}
where $L_z$ is the height of the volume over which to perform the vertical integration.

\begin{figure*}
  \centering
  \noindent\includegraphics[width=41pc]{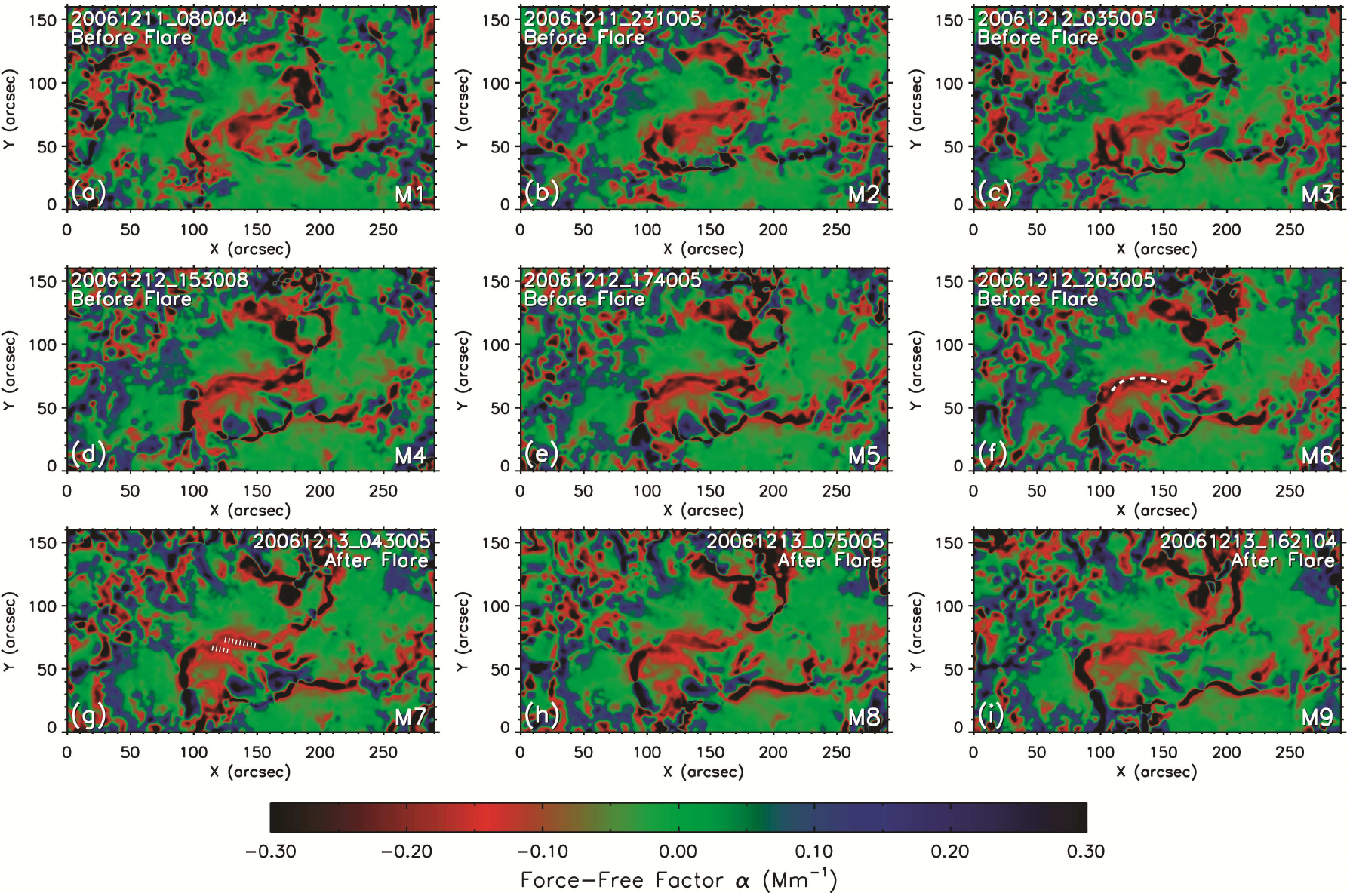}
  \caption{Distribution images of the force-free factor $\alpha$ on layer 2 for the nine reconstructed coronal magnetic field data sets associated with the X3.4 flare event of AR 10930. The data of $\alpha$ are calculated by using equation (\ref{eq:cal-3D-alpha}), and the images are displayed in full FOV. In the marginal area of the active region with weak magnetic field ($|\textbf{B}|$ approaching zero), the values of $\alpha$ are full of noises and may not be reliable. The thick dashed line in Figure 7f and the thick dotted lines in Figure 7g outline the configurations of the local magnetic connectivities around the main PIL before and after the flare as revealed by the distributions of the strong $\alpha$ values. (Please enlarge electronic version of this figure to see detail.)}
\end{figure*}

The maps of $I(x, y)$ for the nine reconstructed coronal magnetic field data sets were computed according to equation (\ref{equ:averaged-current-density}) over the volume between $z=0$ and $z=42$ Mm (i.e., $L_z=42$ Mm). The resulting images are shown in Figure 8 in full FOV. The current density magnitude in Figure 8 is measured in electromagnetic cgs units \citep{Aschwanden_Book_2005} as expressed in equation (\ref{equ:current-density}). The time variation of the current density distributions before and after the X3.4 flare of AR 10930 can be seen in Figure 8 as a reflection of the internal variation of the coronal magnetic fields.

The vertical distributions of current density magnitude in AR 10930 are illustrated and compared with the vertical distributions of force-free factor $\alpha$ in Appendix~A. As seen in Figures A1e and A1f, the electric current density values also present rapid decrease with increase in height.

\subsubsection{Change of Magnetic Connectivity Before and After the Flare as Revealed by the $\alpha$ and Current Density Distribution}
In the $\alpha$ distribution images in Figure 7, it can be seen that there are segments of strong $\alpha$ values (represented by dark red color, $|\alpha|$ greater than $0.15$ Mm$^{-1}$) which extend roughly along the PILs. Considering $\alpha$ constant along each field line, the segment distributions of large $\alpha$ value along PILs should represent local magnetic connectivities in the low corona (except the marginal area of the active region where the $\alpha$ values are noisy and may not be reliable). By comparing with the current density distribution images in Figure 8, it can be seen that in the strong magnetic field areas (see Figure 3), the local magnetic connectivities with strong $\alpha$ values also correspond to the local distributions of strong current density (see section 3.3.4 for further analysis of the relation between $\alpha$ and current density). Hereafter, we refer to this kind of local magnetic connectivity as strong $\alpha$ and strong current density magnetic connectivity, or strong magnetic connectivity for short.

By examining the $\alpha$ and current density distributions before and after the flare (see Figures 7 and 8), it can be found that both the $\alpha$ distributions and the current density distributions associated with M6 and M7 (observed just before and after the flare eruption) show clear differences. There exists a strong magnetic connectivity in the central area of the active region before the flare as shown in Figures 7f and 8f ($\alpha$ and current density distribution images corresponding to M6), which extends along the main PIL (where the X3.4 flare happens) and is outlined by a thick dashed line in Figure 7f. This strong magnetic connectivity is found to be totally broken after the flare eruption as shown in Figures 7g and 8g ($\alpha$ and current density distribution images corresponding to M7). The configuration of the broken magnetic connectivity after the flare is outlined by two thick dotted lines in Figure 7g (corresponding to the two branches of the broken magnetic connectivity). Hereafter, the local region between the two branches of the broken magnetic connectivity is named the magnetic connectivity breaking site or breaking site for short.

The enlarged $\alpha$ (on layer 2) and current density distribution images as well as the $B_z$ images associated with M6 and M7 in the central area of the active region (around the main PIL) are shown in Figure 9. (More $\alpha$ distribution images associated with M6 and M7 on layers of different heights can be found in Appendix~A.) The three images in the left column of Figure 9 (from top to bottom in the order of $B_z$, current density, and $\alpha$) are preflare images (corresponding to M6), and the three images in the right column of Figure 9 are postflare images (corresponding to M7). An isogram of the $\alpha$ distribution (white contour lines) associated with M6 (preflare map) is plotted overlying all the images in Figure 9 to demonstrate the change of magnetic connectivity before and after the flare more clearly. An isogram of the current density distribution associated with M7 (postflare map) is plotted in black color overlying the $\alpha$ isograms in Figures 9b and 9d to show the breaking site of the magnetic connectivity explicitly. The breaking site is also indicated by the thick white arrows in Figures 9b, 9d, and 9f. The directions of the magnetic fields along the magnetic connectivities before and after the flare are determined by comparing with the field line configurations as well as the vector magnetic field distribution on layer 2 of the 3-D coronal magnetic field data (see Appendix~B for details), and are shown in Figures 9e and 9f via thin white arrows.

\begin{figure*}
  \centering
  \noindent\includegraphics[width=41pc]{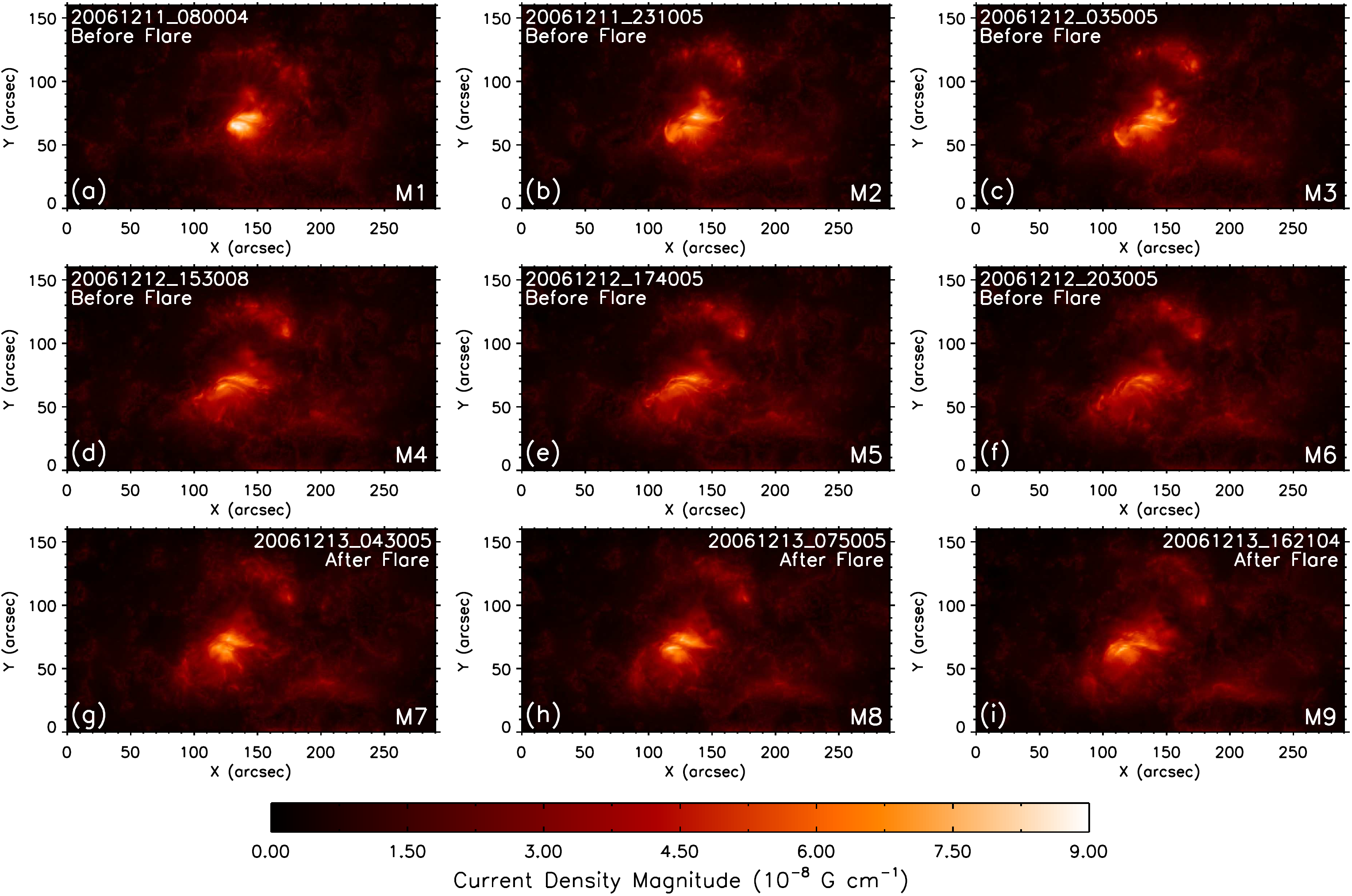}
  \caption{Images of vertically averaged current density intensity map, $I(x, y)$, for the nine sets of reconstructed 3-D coronal magnetic field associated with the X3.4 flare event of AR 10930. The data of $I(x, y)$ are computed by using equation (\ref{equ:averaged-current-density}) over the volume between $z=0$ and $z=42$ Mm (below layer 59 of the 3-D coronal data grid), and the images of $I(x, y)$ are displayed in full FOV. The current density magnitude is measured in electromagnetic cgs units \citep{Aschwanden_Book_2005} as expressed in equation~(\ref{equ:current-density}). (Please enlarge electronic version of this figure to see detail.)}
\end{figure*}

It can be seen in Figure 9b that the breaking site of the strong magnetic connectivity is just on top of a local region featured by the narrow negative flux patch that appears in M7 (see section 3.1 and Figures 5b and 5d for description and clear images of the negative flux patch).

\subsubsection{Further Analysis of the Relation Between $\alpha$ and Current Density}
The distribution images of force-free factor $\alpha$ and current density have similar appearance to some extent (see Figures 7 and 8, and vertical distribution images in Appendix~A) because the two physical measures are related by equation
\begin{linenomath*}
\begin{equation}\label{equ:current-alpha-relation}
  \textbf{j}=(1/4\pi)\alpha\textbf{B}
\end{equation}
\end{linenomath*}
(in the context of NLFFF model), which can be obtained from equations (\ref{equ:current-density}) and (\ref{equ:force-free-alpha}).

As shown in equation (\ref{equ:current-alpha-relation}), the magnitude of the current density is decided by both the values of $\alpha$ and the strength of $\textbf{B}$. Then the distribution of current density can indicates the key areas with both strong $\alpha$ values (representing nonpotential property) and strong magnetic field values (containing more magnetic energy) in active regions (see Figure 8 for example), which is very useful for solar active region and flare forecasting studies \citep{SchrijverEA_ApJ_2008, LekaBarnes_ApJ_2007}. The force-free factor $\alpha$ itself does not reflect the strength of $\textbf{B}$ as implied by equations (\ref{equ:force-free-alpha}) and (\ref{eq:cal-3D-alpha}). Even in the regions with low magnetic field strength, the values of $\alpha$ can be strong (except the noisy data) as demonstrated in Figure 7.

From Figure 9 it can be seen that, around the breaking site of the magnetic connectivity, the value of $\alpha$ after the flare (in Figure 9f, associated with M7) shows a little decrease compared with the value of $\alpha$ before the flare (in Figure 9e, associated with M6), which can be understood as the relaxation of the local magnetic twist \citep{Seehafer_SolPhys_1990}. However, the magnitude of the current density after the flare (see Figure 9d) shows increase compared with the current density magnitude before the flare (see Figure 9c). By using equation (\ref{equ:current-alpha-relation}), the different variation tendencies of the $\alpha$ and current density values can be interpreted by an enhancement of the magnetic field strength in the local region. Although the value of $\alpha$ decreases, the overall effect of $\alpha$ and $\textbf{B}$ in equation (\ref{equ:current-alpha-relation}) is increased, which leads to the magnitude increment of the current density after the flare. (The strength enhancement of the magnetic field around the breaking site of the magnetic connectivity is also supported by the increase of the magnetic energy density in the local region, see section 3.4.2 for details.)

\begin{figure*}
  \centering
  \noindent\includegraphics[width=28.5pc]{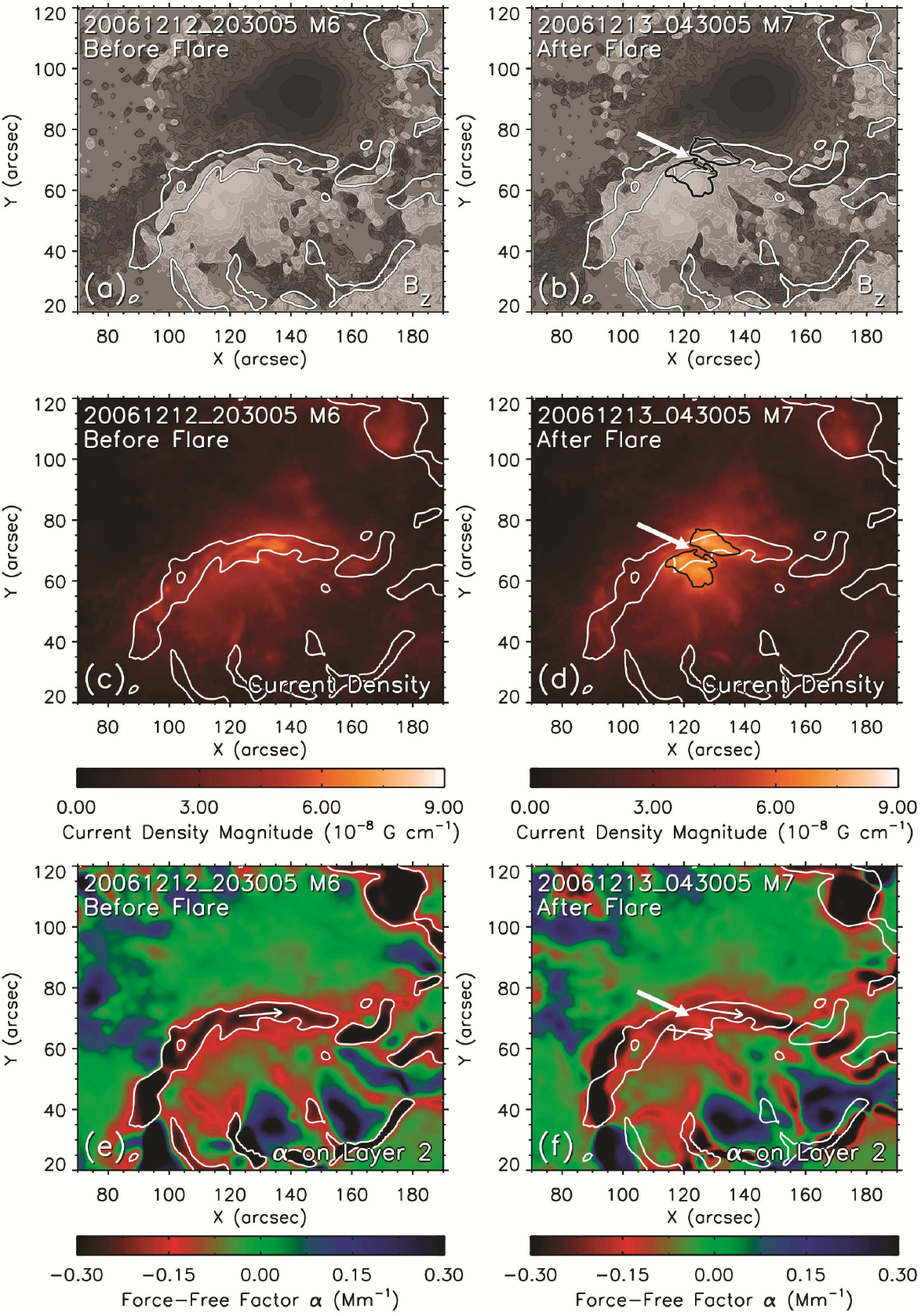}
  \caption{Diagram illustrating the change of magnetic connectivity before and after the X3.4 flare of AR 10930 in the central area of the active region. Images before the flare correspond to M6, and images after the flare correspond to M7. (a and b) $B_z$ contour images of M6 and M7. The contour levels for $B_z$ are $\pm$50, 100, 300, 500, 1000, 1500, 2000, and 3000 G. (c and d) Current density distribution images before and after the flare. (e and f) Distribution images of $\alpha$ on layer 2 of the coronal data grid before and after the flare. An isogram of $\alpha$ (white contour lines, contour level, $-0.18$ Mm$^{-1}$) associated with M6 (preflare map) is plotted overlying all the images to show the location of the strong magnetic connectivity around the main PIL before the flare (see main text for the explanation of the strong magnetic connectivity). A current density isogram (contour level, $6.10 \times 10^{-8}$ G cm$^{-1}$) associated with M7 (postflare map) is plotted overlying the $\alpha$ isograms in Figures 9b and 9d in black color to show the breaking site of the magnetic connectivity after the flare. The breaking site is also indicated by the thick white arrows in Figures 9b, 9d, and 9f. The thin white arrows in Figures 9e and 9f indicate the directions of the magnetic fields along the magnetic connectivities before and after the flare.}
\end{figure*}

\subsection{Variation of the Magnetic Energy}
The magnetic energy of coronal magnetic field contained in a volume $\Omega$ can be expressed by equation \citep{Sakurai_SSRv_1989, RegnierPriest_ApJ_2007}:
\begin{linenomath*}
\begin{equation}\label{eq:magnetic-energy}
  E = \int_{\Omega}\frac{B^2}{8\pi}\textrm{d}\Omega.
\end{equation}
\end{linenomath*}
By restricting the $\Omega$ to the scale of the volume unit, the equation (\ref{eq:magnetic-energy}) can also be used for calculating the magnetic energy density values.

In this subsection, we use the reconstructed coronal magnetic fields of NLFFF model to evaluate the magnetic energy stored in AR 10930 and analyze the time variation of the magnetic energy density spatial distributions before and after the X3.4 flare.

\subsubsection{Total Magnetic Energy}
The total magnetic energy values associated with the nine sets of reconstructed 3-D coronal magnetic fields (represented by $E_{\mathrm{NLFFF}}$) were calculated over the $290 \times 160 \times 160$ pixel volume ($1''$/pixel, see section 2.3) by using equation (\ref{eq:magnetic-energy}), and the results are listed in Table 2. The time series variation of the total energy values is illustrated in Figure 10 through square symbols, with the solar soft X-ray flux curve as the background for reference. As shown in Table 2 and Figure 10, the magnitude of the total magnetic energy ($E_{\mathrm{NLFFF}}$) in AR 10930 has the order of $10^{33}$ erg and the energy fluctuation range has the order of $10^{32}$ erg, which is consistent with the magnetic energy values of AR 10930 in the literatures \citep[e.g.,][]{SchrijverEA_ApJ_2008}.

The magnetic energy of the potential field (extrapolated from the same magnetogram data as used by the NLFFF extrapolation, but only utilizing the $B_z$ component), $E_{\mathrm{PF}}$, and the free magnetic energy, $E_{\mathrm{free}}$ (defined as $E_{\mathrm{free}}=E_{\mathrm{NLFFF}}-E_{\mathrm{PF}}$) \citep{RegnierPriest_ApJ_2007}, were also calculated for the nine coronal data sets. The free magnetic energy gives an estimate of the available portion of energy that powers the solar explosive events. The results of $E_{\mathrm{PF}}$ and $E_{\mathrm{free}}$ are listed in Table 2 with the values of $E_{\mathrm{NLFFF}}$, and are plotted in Figure 10 by using diamond symbols (for $E_{\mathrm{PF}}$) and triangle symbols (for $E_{\mathrm{free}}$), respectively. The values of the total unsigned magnetic flux through the photosphere (represented by $\Phi$), which is related with the values of magnetic energy in the corona, are included in Table 2 and plotted in Figure 10 (through dotted line and asterisk symbols) as well for reference.

\begin{table*}
  \caption{Evaluation of Total Magnetic Energy and Total Unsigned Magnetic Flux for the Nine Sets of Reconstructed 3-D Coronal Magnetic Field Associated With the X3.4 Flare Event of AR 10930\tablenotemark{a}}
  \centering
  \begin{tabular}{c c c c c c}
  \hline
  & Total Magnetic  & & & Total Unsigned  &  \\
  &  Energy $E_{\mathrm{NLFFF}}$\tablenotemark{c}  & $E_{\mathrm{PF}}$\tablenotemark{d}  & $E_{\mathrm{free}}$\tablenotemark{e}  & Magnetic Flux $\Phi$\tablenotemark{f}  & Label of  \\
  Time of Observation\tablenotemark{b}  &   ($10^{33}$ erg)  & ($10^{33}$ erg) & ($10^{33}$ erg) & ($10^{22}$ Mx) & Magnetogram  \\
  \hline
  2006-12-11 08:00:04 UT  &  4.329  &  2.285  & 2.044  & 5.784 &  M1 \\
  2006-12-11 23:10:05 UT  &  4.401  &  2.305  & 2.096  & 5.607 &  M2 \\
  2006-12-12 03:50:05 UT  &  4.461  &  2.287  & 2.174  & 5.603 &  M3 \\
  2006-12-12 15:30:08 UT  &  4.630  &  2.322  & 2.308  & 5.853 &  M4 \\
  2006-12-12 17:40:05 UT  &  4.529  &  2.363  & 2.166  & 5.914 &  M5 \\
  2006-12-12 20:30:05 UT  &  4.308  &  2.343  & 1.965  & 5.867 &  M6 \\
  2006-12-13 04:30:05 UT  &  4.191  &  2.300  & 1.891  & 5.850 &  M7 \\
  2006-12-13 07:50:05 UT  &  4.194  &  2.313  & 1.881  & 5.814 &  M8 \\
  2006-12-13 16:21:04 UT  &  4.267  &  2.334  & 1.933  & 5.901 &  M9 \\
  \hline
  \end{tabular}
  \tablenotetext{a}{The magnetic energy values were calculated over the $290 \times 160 \times 160$ pixel volume of the 3-D coronal data grid ($1''$/pixel). The time series curves of the magnetic energy and magnetic flux values are shown in Figure 10.}
  \tablenotetext{b}{Dates are formatted as year/month/day.}
  \tablenotetext{c}{$E_{\mathrm{NLFFF}}$ is the total magnetic energy contained in the reconstructed 3-D coronal magnetic fields of NLFFF model. The values of $E_{\mathrm{NLFFF}}$ were calculated by using equation (\ref{eq:magnetic-energy}).}
  \tablenotetext{d}{$E_{\mathrm{PF}}$ is the magnetic energy of the potential field. The potential field data are extrapolated from the same magnetogram data as used by the NLFFF extrapolation, but only utilizing the $B_z$ component. Fourier transform approach \citep{Sakurai_SSRv_1989} was employed for the potential field extrapolations. The values of $E_{\mathrm{PF}}$ were calculated by using equation (\ref{eq:magnetic-energy}) and based on the potential field data.}
  \tablenotetext{e}{$E_{\mathrm{free}}$ is free magnetic energy, which is defined as $E_{\mathrm{free}}=E_{\mathrm{NLFFF}}-E_{\mathrm{PF}}$.}
  \tablenotetext{f}{The values of the total unsigned magnetic flux through the photosphere (represented by $\Phi$) were calculated from the $B_z$ components of the magnetogram data used by the NLFFF extrapolations.}
\end{table*}

\begin{figure}[htpd]
  \centering
  \noindent\includegraphics[width=22pc]{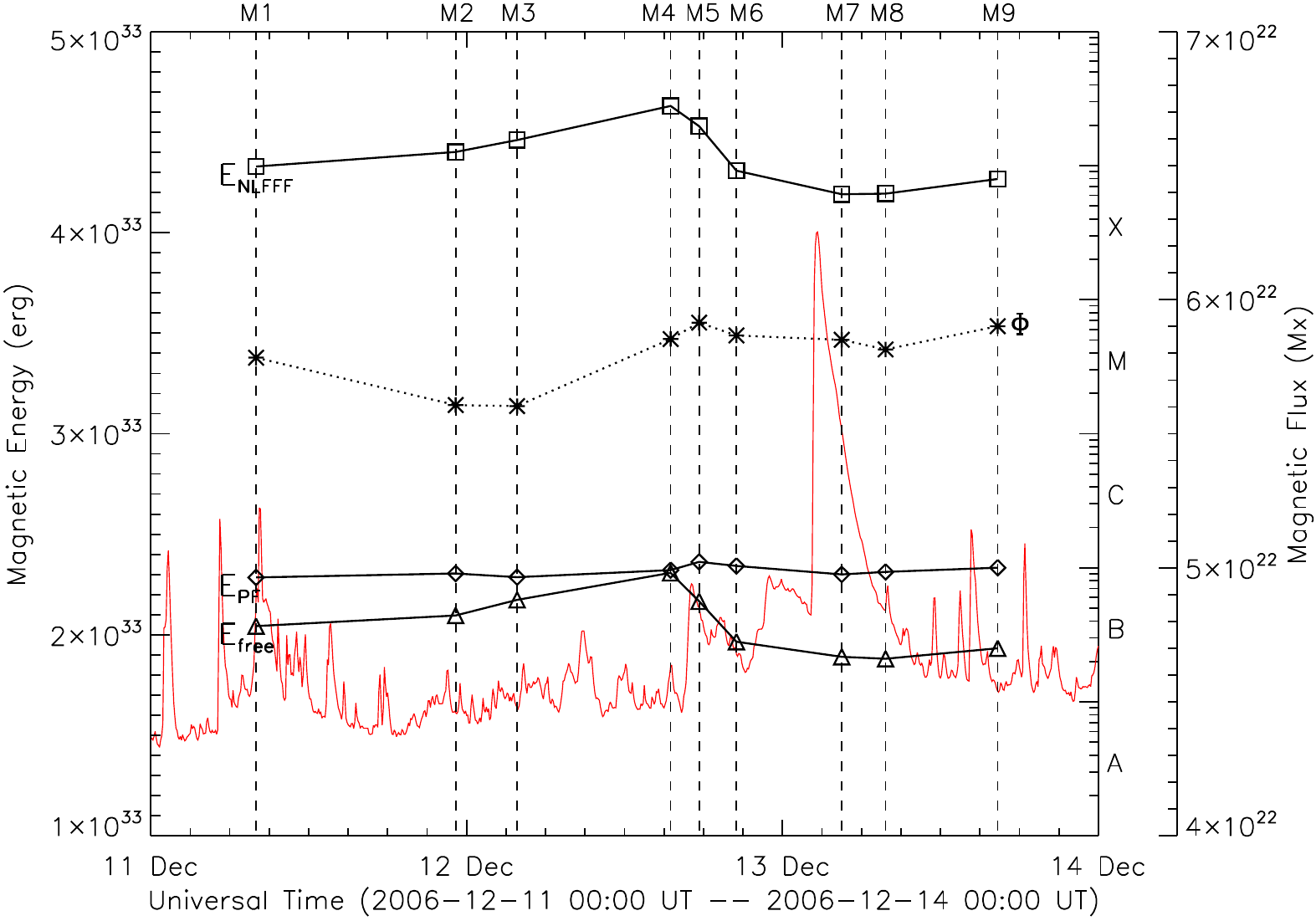}
  \caption{Time series curves of the $E_{\mathrm{NLFFF}}$, $E_{\mathrm{PF}}$, $E_{\mathrm{free}}$, and $\Phi$ values calculated from the nine sets of reconstructed 3-D coronal magnetic field data associated with the X3.4 flare event of AR 10930. $E_{\mathrm{NLFFF}}$ is the total magnetic energy contained in the reconstructed 3-D coronal magnetic fields of NLFFF model (plotted using square symbols). $E_{\mathrm{PF}}$ is the magnetic energy of the corresponding potential field (plotted using diamond symbols). $E_{\mathrm{free}}$ is defined as $E_{\mathrm{free}}=E_{\mathrm{NLFFF}}-E_{\mathrm{PF}}$ (plotted using triangle symbols). $\Phi$ is total unsigned magnetic flux through the photosphere (plotted using dotted line and asterisk symbols). The solar soft X-ray flux curve (obtained by the GOES 11 satellite in 1.0--8.0~{\AA} wavelength band) is displayed as the background for reference. The vertical dashed lines indicate the observation times of the nine Hinode/SOT-SP magnetograms selected for coronal magnetic field reconstruction, which are marked from M1 to M9, respectively. The exact values of $E_{\mathrm{NLFFF}}$, $E_{\mathrm{PF}}$, $E_{\mathrm{free}}$, and $\Phi$ can be found in Table 2.}
\end{figure}

It can be seen in Table 2 and Figure 10 that the fluctuation range of the potential field energy ($E_{\mathrm{PF}}$) has the order of $10^{31}$ erg, and the fluctuation range of the free magnetic energy ($E_{\mathrm{free}}$) has the order of $10^{32}$ erg (the same order as the fluctuation range of $E_{\mathrm{NLFFF}}$). Then the profile of the $E_{\mathrm{free}}$ curve in Figure 10 is dominated by the $E_{\mathrm{NLFFF}}$ variation. Since free magnetic energy is a global concept and $E_{\mathrm{NLFFF}}$ dominates the profile of the free magnetic energy variation in AR 10930, our following analysis of the time variation of the magnetic energy spatial distribution before and after the flare is based on the 3-D magnetic energy density data associated with $E_{\mathrm{NLFFF}}$.

\subsubsection{Variation of the Magnetic Energy Spatial Distribution Before and After the Flare}
In order to find out the changes of magnetic energy spatial distribution before and after the X3.4 flare of AR 10930, we first calculate the magnetic energy contained in each layer volume (one-pixel-high volume beneath each layer, counting from layer 1 \citep{HeEA_JGR_2011}) for the two sets of reconstructed 3-D coronal magnetic fields associated with M6 and M7 (observed just before and after the flare eruption), and then produce the difference curve of the two sets of magnetic energy vertical distributions (i.e., $E_{M7}-E_{M6}$). The resulting plot is displayed in Figure 11a.

\begin{figure*}
  \centering
  \noindent\includegraphics[width=32pc]{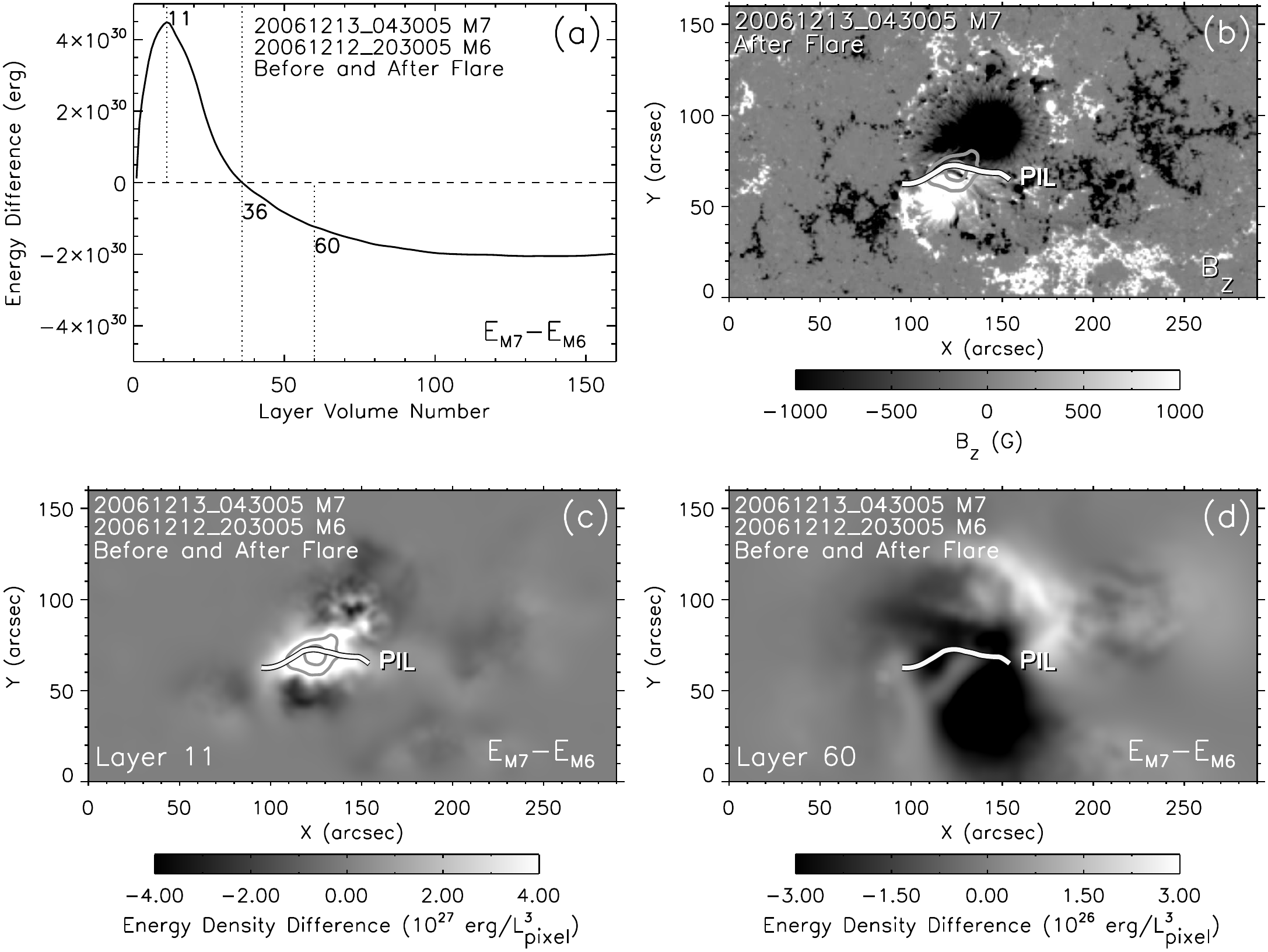}
  \caption{Diagram illustrating the locations at which the coronal magnetic energy density in AR 10930 varies before and after the X3.4 flare eruption. (a) Difference curve of the magnetic energy vertical distributions between $E_{M6}$ and $E_{M7}$. The three vertical dotted lines (from left to right) denote the layer 11 (maximum energy increment layer), layer 36 (separation layer between energy increasing and decreasing), and layer 60 (representative layer of energy decreasing), respectively. (b) $B_z$ image of M7 (observed just after the flare eruption). The main PIL between the main negative and positive polarities in the active region is indicated by a thick white line, which is also drawn in Figures 11c and 11d. (c and d) Difference images of the energy density horizontal distributions between $E_{M6}$ and $E_{M7}$ at layer 11 and layer 60, respectively. The volume unit used for calculating the energy density is a cube with one pixel length ($L_{\textrm{pixel}}$) in $X$, $Y$, and $Z$ direction. In Figure 11c, the distribution of the energy density difference ($E_{M7}-E_{M6}$) inside the saturation area is displayed by two contour lines (contour levels, $6.0 \times 10^{27}, 12.0 \times 10^{27} \mathrm{erg}/L_{\mathrm{pixel}}^{3}$), which are also plotted in Figure 11b.}
\end{figure*}

From the profile of the curve in Figure 11a (vertical distribution of the magnetic energy difference before and after the flare), it can be seen that in the lower space of the modeling volume (below layer 36), the magnetic energy increases, and the maximum increment is at layer 11, while in the higher space (above layer 36), the energy decreases. To find out the locations at which the energy density varies, we make the difference images of the energy density horizontal distributions at layer 11 (maximum energy increment layer) and layer 60 (representative layer of energy decreasing), respectively. The two energy density difference images together with the $B_z$ image of M7 are also displayed in Figure 11. The main PIL of M7 is indicated in the $B_z$ image of M7 (Figure 11b) by a thick white line, which is also plotted on the energy density difference images (Figures 11c and 11d) to show the positions of energy density increasing and decreasing more clearly.

As demonstrated in Figures 11c and 11d, the region of magnetic energy density increasing within the lower layers is located just above the main PIL of the photospheric magnetogram (see Figure 11c), and the region of energy density decreasing in the higher space is also above the main PIL, though with different shape and orientation (see Figure 11d). The position of the main PIL is where the X3.4 flare of AR 10930 happened \citep{SchrijverEA_ApJ_2008}, thus the variations of the magnetic energy spatial distributions shown in Figure 11 are directly related with the flare eruption.

In Figure 11c, the gray scale for the energy density difference map at layer 11 saturates at $4.0 \times 10^{27} \mathrm{erg}/L_{\mathrm{pixel}}^{3}$ ($L_{\mathrm{pixel}}$ representing the length of 1 pixel). Inside the saturation area, the distribution of the energy density difference is illustrated by two contour lines (contour levels, $6.0 \times 10^{27}, 12.0 \times 10^{27} \mathrm{erg}/L_{\mathrm{pixel}}^{3}$), and the inner contour line encloses the region with the highest values of energy density increase. The two contour lines are also plotted in Figure 11b ($B_z$ image of M7).

As shown in Figure 11b, the site with the highest values of energy density increase at layer 11 is just above a small area around the narrow negative flux patch in the photospheric magnetogram M7 (see section 3.1 and Figures 5b and 5d for description and clear images of the negative flux patch).

\section{Comparison With the Flare Onset Imaging Observation}
The analyses of the 3-D distributions of the coronal physical measures in section 3 for the nine sets of reconstructed coronal magnetic field data associated with the X3.4 flare of AR 10930 manifest the distinct variations of the coronal magnetic fields before and after the flare. In this section, we compare the distribution variations of the coronal physical measures before and after the flare as shown in Figures 9 and 11 with the flare onset imaging observation to investigate the spatial relationship between the site of the distinct variations of the coronal physical measures and the location of the flare initial eruption.

\begin{figure*}
  \centering
  \noindent\includegraphics[width=28pc]{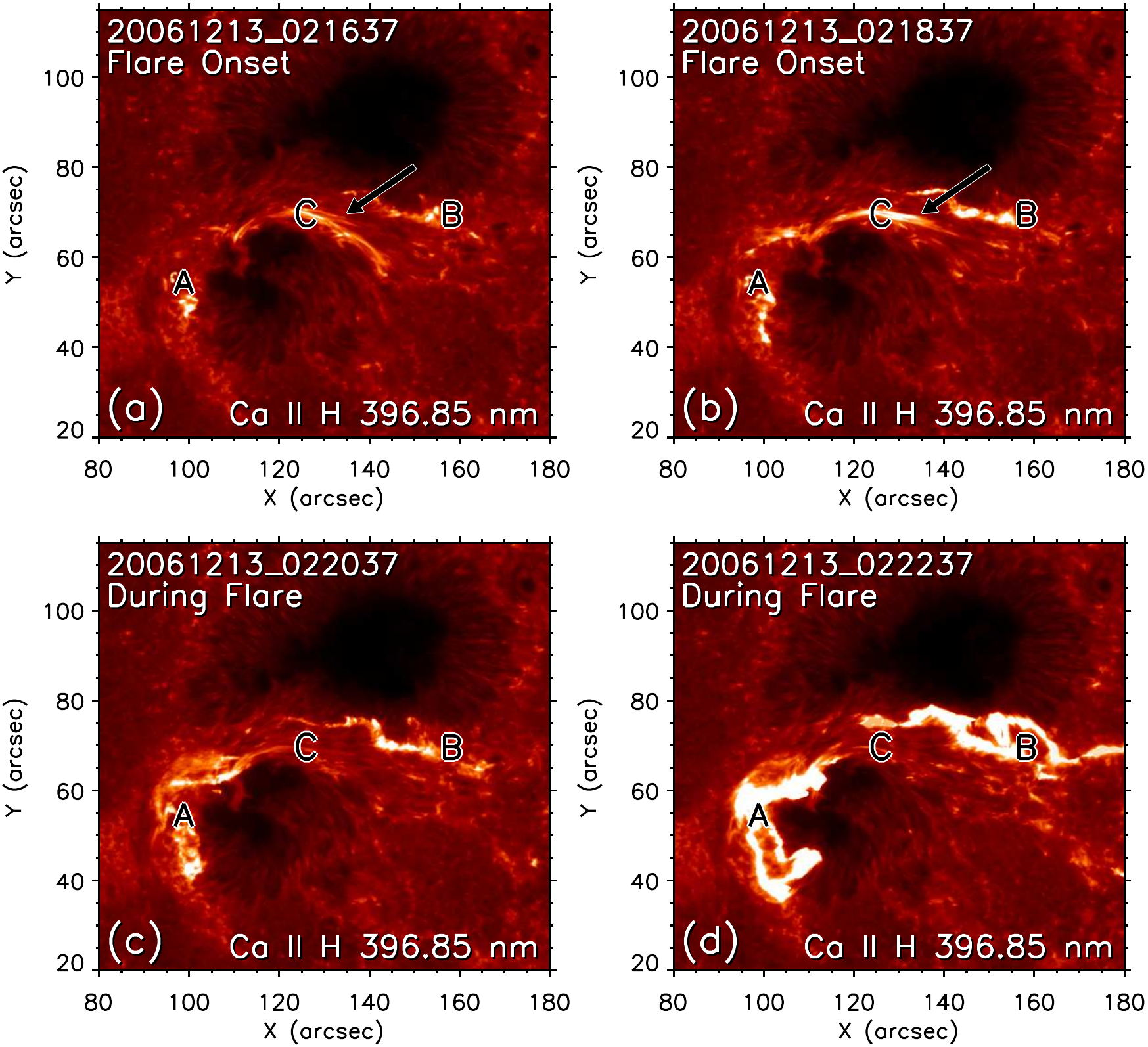}
  \caption{Four Ca \textsc{ii} H line (396.85 nm) filtergrams observed during the onset and rising phase of the X3.4 flare of AR 10930 in the central area of the active region. The data of the four filtergrams were acquired by Hinode/SOT at 13 December 2006 02:16:37 UT (2 min after the flare start), 02:18:37 UT, 02:20:37 UT, and 02:22:37 UT, respectively. The images of the four filtergrams have been corrected for the projection effect and are coaligned with the Hinode/SOT-SP magnetograms. The capital letters A and B in the images denote the two flare ribbons. The thick black arrows in Figures 12a and 12b indicate the initial eruptive structures during the flare onset phase, which manifest as the thin straight fibres in the filtergrams. The capital letter C in the images denotes the location of the flare initial eruption as inferred from the initial eruptive structures and the two flare ribbons.}
\end{figure*}

The flare images selected for the comparison are four Ca \textsc{ii} H line (396.85 nm) filtergrams of flare onset and rising phase observed by Hinode/SOT \citep{TsunetaEA_SolPhys_2008} at 13 December 2006 02:16:37 UT (2 min after the flare start), 02:18:37 UT, 02:20:37 UT, and 02:22:37 UT, respectively, in which the initial eruptive structures during the flare onset phase and the two flare ribbons can be identified clearly. We perform the similar procedure of projection effect correction as in section 2.2 on the Ca \textsc{ii} H images to achieve the co-alignment with the magnetograms. The four projection-effect-corrected and aligned Ca \textsc{ii} H images in the central area of the active region are shown in Figure 12. The two flare ribbons are denoted by the capital letters A and B, respectively, for all the images in Figure 12. The initial eruptive structures, which manifest as the thin straight fibres during the flare onset phase, are indicated by the thick black arrows in Figures 12a and 12b. The capital letter C in Figure 12 denotes the location of the flare initial eruption as inferred from the eruptive structures during the flare onset phase and the two flare ribbons.

In Figure 13, we compare the Ca \textsc{ii} H image observed on 13 December 2006 at 02:16:37 UT (as Figure 13a, flare onset phase, also shown in Figure 12a) with the current density distribution image associated with M7 (as Figure 13d, after flare), the $\alpha$ distribution image on layer 2 associated with M7 (as Figure 13e, after flare), and the energy density difference image at layer 11 between $E_{M6}$ and $E_{M7}$ (as Figure 13f, before and after flare) in the central area of the active region. The $B_z$ contour image and gray scale image of M7 (observed just after the flare eruption) are included in Figure 13 (as Figures 13b and 13c) for reference. The isogram of the current density distribution associated with M7 shown in Figure 9d (drawn in black color) and the contour lines of the energy density difference map ($E_{M7}-E_{M6}$ at layer 11) shown in Figure 11c are plotted in Figures 13a and 13b and 13d--13f to demonstrate the spatial correlations between the features of the individual images. The thick white arrows in Figures 13a--13c indicate the breaking site of the magnetic connectivity revealed by the distribution variations of $\alpha$ and current density before and after the flare as shown in Figure 9. The capital letter C in Figures 13a and 13c indicates the location of the flare initial eruption inferred from the Ca \textsc{ii} H images as shown in Figure 12.

\begin{figure*}
  \centering
  \noindent\includegraphics[width=40pc]{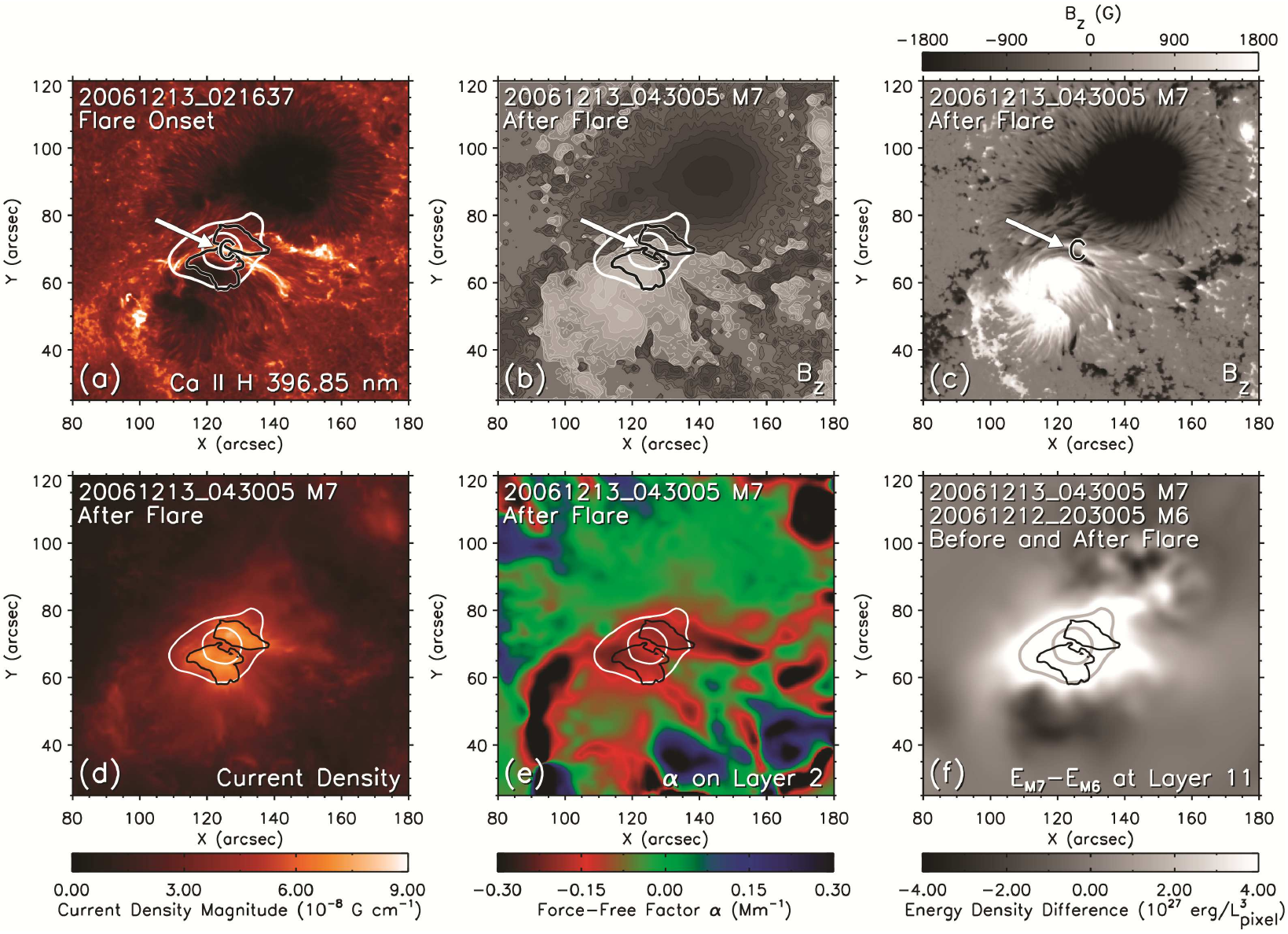}
  \caption{Comparison between the distribution variations of the coronal physical measures before and after the X3.4 flare of AR 10930 and the flare onset imaging observation in the central area of the active region. (a) Ca \textsc{ii} H line (396.85 nm) filtergram of flare onset phase observed by Hinode/SOT at 13 December 2006 02:16:37 UT (2 min after the flare start). The Ca \textsc{ii} H image has been corrected for the projection effect and is co-aligned with other images (see also Figure 12a). (b and c) $B_z$ contour image and gray scale images of M7 (observed just after the flare eruption), which are included for reference. The contour image of $B_z$ is based on the rebinned data ($1''$/pixel) of M7, and the contour levels are $\pm$50, 100, 300, 500, 1000, 1500, 2000, and 3000 G. The gray scale image of $B_z$ is based on the full resolution data ($0.25''$/pixel) of M7, and the gray scale saturates at $\pm 1800$ G. (d) Current density distribution image associated with M7 (see also Figure 9d). (e) The $\alpha$ distribution image on layer 2 associated with M7 (see also Figure 9f). (f) Energy density difference image at layer 11 between $E_{M6}$ and $E_{M7}$ (see also Figure 11c). The isogram of the current density distribution shown in Figure 9d (contour level, $6.10 \times 10^{-8}$ G~cm$^{-1}$, drawn in black color) and the contour lines of the energy density difference map shown in Figure 11c (contour levels, $6.0 \times 10^{27}, 12.0 \times 10^{27} \mathrm{erg}/L_{\mathrm{pixel}}^{3}$) are also plotted in Figures 13a and 13b and 13d--13f to demonstrate the spatial correlations between the features of the individual images. The thick white arrows in Figures 13a--13c indicate the breaking site of the strong magnetic connectivity as shown in Figure 9. The capital letter C in Figures 13a and 13c indicates the location of the flare initial eruption as shown in Figure 12.}
\end{figure*}

In Figure 13a, it can be seen that both the breaking site of magnetic connectivity (indicated by the thick white arrow and the isogram of current density in Figure 13a) and the site with the highest values of energy density increase before and after the flare ($E_{M7}-E_{M6}$, indicated by the inner contour line of the energy density difference map in Figure 13a) coincide with the location of the flare initial eruption (indicated by the capital letter C in Figure 13a). In Figure 13c, it can be seen that the location of the flare initial eruption (indicated by the thick white arrow and the capital letter C in Figure 13c) is just above the north edge of the narrow negative flux patch in the photospheric magnetogram M7 (see section 3.1 and Figures 5b and 5d for description and more clear images of the negative flux patch).

\section{Discussion}
\subsection{What Causes the Broken Magnetic Connectivity}
The distribution variations of $\alpha$ and current density shown in section 3.3 reveal a distinct change of magnetic connectivity in the low corona of AR 10930 during the X3.4 flare. There exists a strong $\alpha$ and strong current density magnetic connectivity along the main PIL of the active region before the flare, and the strong magnetic connectivity is totally broken after the flare eruption (see Figure 9). Since the variations of the coronal magnetic field are the responses to the variations of the photospheric vector magnetic field \citep{HeEA_JGR_2011}, this breaking of magnetic connectivity in the low corona should be driven by the magnetic field changes in the photosphere.

As shown in Figure 9b (see also Figures 13b and 13c), the breaking site of the strong magnetic connectivity is just over a local area featured by the narrow negative flux patch that appears in the photospheric magnetogram M7 (see section 3.1 for description of the negative flux patch), or more exactly, the north edge of the narrow patch as shown in Figure 13c. Considering the counterclockwise rotation of the main positive polarity of AR 10930 (see Figures 5a and 5c), it can be inferred that it is the appearance of this negative flux patch along with the motion of the main positive polarity in the photosphere that leads to the breaking of the strong magnetic connectivity in the low corona.

\begin{figure*}
  \centering
  \noindent\includegraphics[width=23pc]{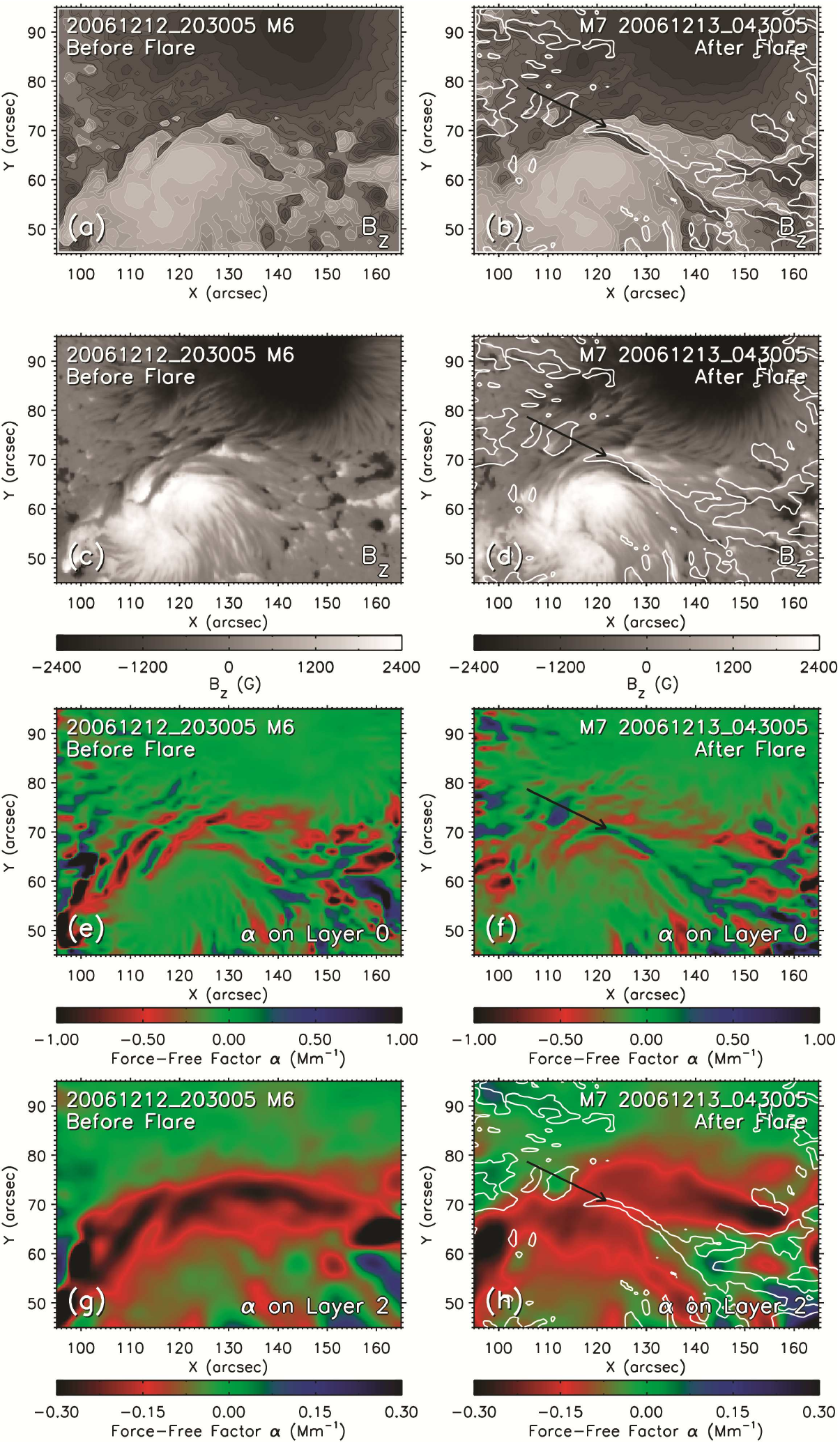}
  \caption{Comparison between the $\alpha$ distributions around the main PIL on layer 0 and layer 2 for the data before and after the X3.4 flare of AR 10930. Images before the flare correspond to M6, and images after the flare correspond to M7. (a--d) $B_z$ contour images and gray scale images of M6 and M7, which are included for reference. The contour images are based on the rebinned data ($1''$/pixel) of the magnetograms, and the contour levels are $\pm$50, 100, 300, 500, 1000, 1500, 2000, and 3000 G. The gray scale images are based on the full resolution data ($0.25''$/pixel), and the gray scale saturates at $\pm 2400$ G. (e and f) $\alpha$ distribution images associated with M6 and M7 on layer 0 (representing the photosphere). (g and h) $\alpha$ distribution images associated with M6 and M7 on layer 2. The thin black arrow in Figures 14f indicate a long cramped region with positive $\alpha$ values (displayed by blue color) in the $\alpha$ image on layer 0 associated with M7. An isogram of the positive $\alpha$ distribution (contour level, $0.06$ Mm$^{-1}$) as well as the thin black arrow in Figure 14f is plotted in Figures 14b, 14d, and 14h to show the location and coverage area of the cramped positive $\alpha$ region on these images.}
\end{figure*}

To verify this conclusion, we compare the $\alpha$ distributions around the main PIL on layer 0 (bottom layer, representing the photosphere) and layer 2 for the data associated with M6 and M7 (observed just before and after the flare eruption). The resulting images are shown in Figure 14 (third and fourth row). The $B_z$ contour images and gray scale images of M6 and M7 are also included in Figure 14 (first and second row) for reference. From the $\alpha$ image on layer 0 associated with M7 (Figure 14f), it can be found that in the central area of the image, there exists a long cramped region (indicated by a thin black arrow) in which the $\alpha$ values are positive (displayed by blue color in Figure 14f). An isogram of the positive $\alpha$ distribution in Figure 14f, which can represent the location and coverage area of the cramped positive $\alpha$ region, is plotted on the $B_z$ images of M7 (Figures 14b and 14d) and the $\alpha$ image on layer 2 associated with M7 (Figure 14h). In Figures 14b and 14d, it can be seen that the cramped region with positive $\alpha$ values is just located at the north edge of the narrow negative flux patch over which the breaking of the magnetic connectivity happens. In Figure 14h, it can be seen that the position of the cramped positive $\alpha$ region coincides with the separation line between the two branches of the broken magnetic connectivity.

Considering its positive $\alpha$ values, the magnetic connectivity property of the cramped region shown in Figure 14f is different from the surrounding area where the negative $\alpha$ values dominate. Along with the counterclockwise rotation of the main positive polarity in the photosphere, this cramped positive $\alpha$ region moves into the field where the strong magnetic connectivity resides before the flare, splits the strong magnetic connectivity during the flare eruption, and causes the broken magnetic connectivity configuration after the flare.

The injection of positive $\alpha$ into the existing negative $\alpha$ region in AR 10930 is consistent with the result obtained by \citet{ParkEA_ApJ_2010} in terms of magnetic helicity and the result obtained by \citet{InoueEA_ApJ_2012} in terms of magnetic twists.

\subsection{Scenario for the Initial Eruption of the Flare}
The spatial coincidence of the breaking site of the strong magnetic connectivity and the location of the flare initial eruption (see Figure 13 and description in section 4) implies that it is the breaking of the magnetic connectivity that induces the initial plasma ejection of the flare. The process can be understood as follows: The breaking of the strong magnetic connectivity let the corona system become nonequilibrium and causes isolated electric current at the breaking site. The isolated electric current is the residual and variant of the strong current (associated with the strong magnetic connectivity) before the flare, which is not compatible with the broken magnetic configuration and thus isolated. To achieve a new equilibrium state, this part of excess (isolated) electric current and the associated plasmas must be expelled from the corona system, then the initial plasma ejection of the flare happens. The ejected plasmas in turn cause further magnetic reconnections and postflare loops \citep{ShibataMagara_LRSP_2011}.

The diagram of this scenario for the flare initial eruption is shown in Figure 15. The configuration of the strong magnetic connectivity before the flare (represented by a thick black line) is sketched in Figure 15a (corresponding to the state associated with M6, see Figure 15b). Note that the current density vector and the magnetic field vector associated with the strong magnetic connectivity have opposite directions since the local distribution of force-free factor $\alpha$ shows apparent negative values (see description in section 3.3.1). The configuration of the broken magnetic connectivity after the flare (represented by two thick dotted lines, see Appendix~B for field lines configurations) is sketched in Figure 15e (corresponding to the state associated with M7, see Figure 15f). The nonequilibrium state of the initial eruption of the flare is sketched in Figure 15c (also shown in Figure 15d over the magnetogram and isogram of $\alpha$ associated with M7).

\begin{figure*}
  \centering
  \noindent\includegraphics[width=30pc]{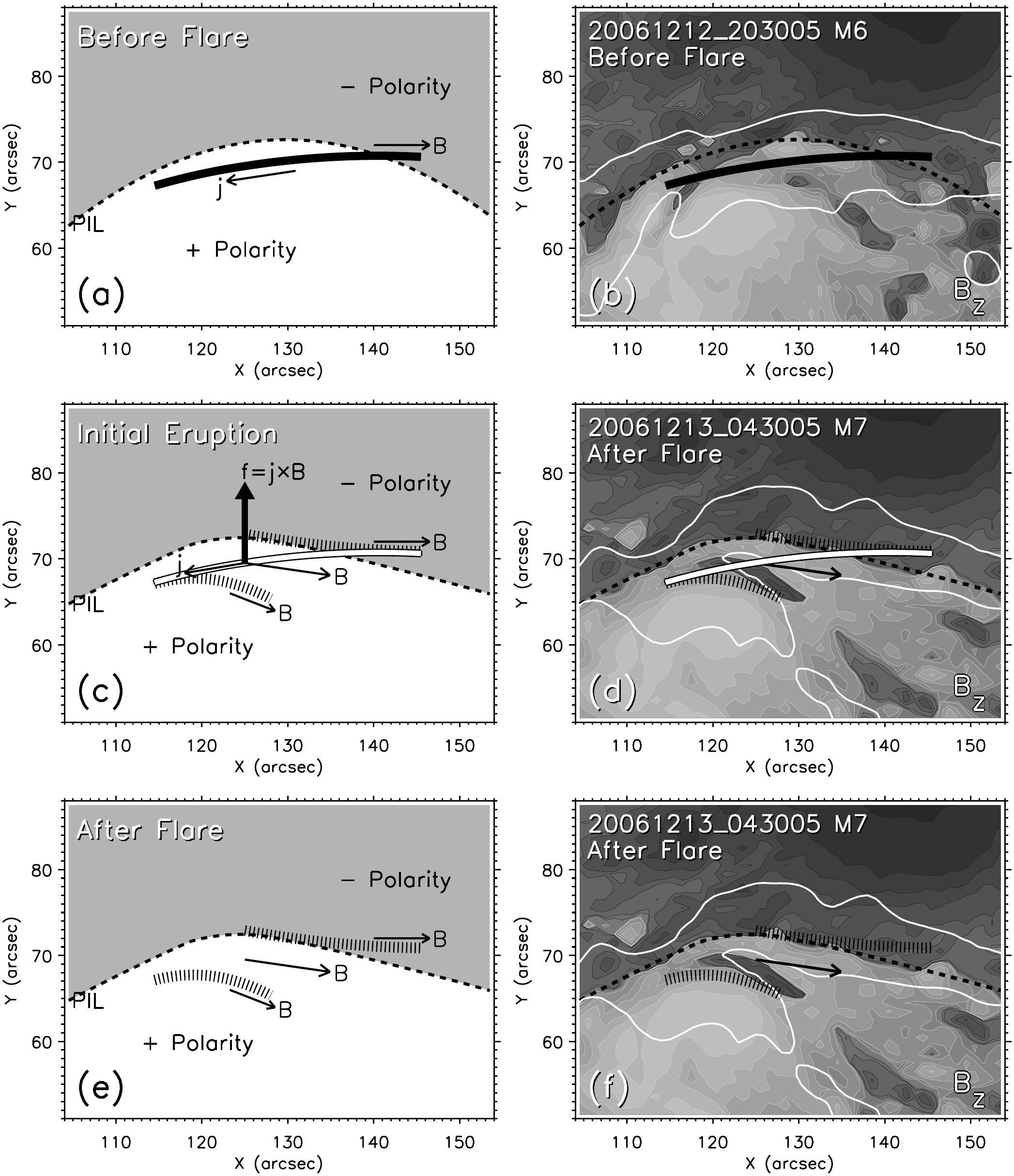}
  \caption{Sketch illustrating the scenario for the initial eruption of the X3.4 flare of AR 10930 induced by the breaking of the strong magnetic connectivity around the main PIL of the active region. (a and b) Equilibrium state of the strong magnetic connectivity before the flare and the corresponding photospheric magnetogram M6 and isogram of $\alpha$ on layer 2. The strong magnetic connectivity is represented by a thick black line. Note that the current density vector and the magnetic field vector associated with the strong magnetic connectivity have opposite directions. (c and d) Nonequilibrium state of the initial eruption of the flare and the corresponding photospheric magnetogram M7 and isogram of $\alpha$ on layer 2. The two thick dotted lines represent the configuration of the broken magnetic connectivity (see Appendix B for field lines configuration). The thick white line represents the isolated electric current at the magnetic connectivity breaking site. The upward thick black arrow represents the Lorentz force acting on the isolated electric current which lifts the associated plasmas and causes the initial plasma ejection. (e and f) Equilibrium state of the broken magnetic connectivity after the flare and the corresponding photospheric magnetogram M7 and isogram of $\alpha$ on layer 2. The thin black arrows in Figures 15d and 15f indicate the representative direction of the magnetic field at the magnetic connectivity breaking site after the flare (see Appendix~B for details). The contour level for the isograms of $\alpha$ in Figures 15b, 15d, and 15f is $-0.15$ Mm$^{-1}$.}
\end{figure*}

As illustrated in Figures 15c and 15d, while the broken configuration of the magnetic connectivity has just been formed, there exists isolated electric current (residual and variant of the strong current before the flare, represented by a thick white line in Figures 15c and 15d) between the two branches of the broken magnetic connectivity, and the orientation of the isolated electric current is not parallel to the direction of the surrounding ambient magnetic field which is represented by a thin black arrow in Figure 15d (associated with the broken magnetic configuration, see Appendix~B for details about the magnetic field direction). Note that Figures 15c and 15d correspond to nonequilibrium state and thus force-free constraint can be violated, then there is Lorentz force at play. As demonstrated in Figure 15c, it is the Lorentz force (represented by an upward thick black arrow) acting on the isolated electric current at the magnetic connectivity breaking site that lifts the associated plasmas and causes the initial plasma ejection. The two branches of the broken magnetic connectivity (the two thick dotted lines in Figure 15c) together with the isolated electric current at the magnetic connectivity breaking site (the thick white line in Figure 15c) show a Z-shaped configuration (see also Figure 15d).

It should be noted that the broken magnetic connectivity is a 3-D structure (see Appendix~A for the vertical distributions of $\alpha$ and current density). Not only the lower core magnetic field at the breaking site but also the overlying magnetic structure is disrupted along with the breaking of the magnetic connectivity. Between the two branches of the broken magnetic connectivity is a localized yet opened magnetic channel (see Figures A1d and A1f). The isolated electric current and the associated plasmoid at the magnetic connectivity breaking site can be ejected along the channel straightforwardly by the Lorentz force as described in the above scenario.

\subsection{Redistribution of Magnetic Energy Before and After the Flare}
The results in section 3.4 show that the total magnetic energy of AR 10930 is always in changing along with the evolution of the 3-D coronal magnetic field. As displayed in Table 2 and Figure 10, the difference between the maximum and minimum values of the total magnetic energy ($E_{\mathrm{NLFFF}}$) is $4.39 \times 10^{32}$ erg ($E_{M4}-E_{M7}$), whereas the total magnetic energy change before and after the flare eruption is $1.17 \times 10^{32}$ erg ($E_{M6}-E_{M7}$), which is only a small fraction of the maximum range of the total energy fluctuation.

By investigating the changes of the magnetic energy spatial distributions before and after the flare, we have a different view of the variation of magnetic energy. The magnetic energy in the modeling volume is not uniformly increasing or decreasing before and after the flare. Instead, in the lower space of the modeling volume the energy increasing dominates, while in the higher space the energy decreasing dominates (see Figure 11). That is, there exists magnetic energy redistribution before and after the flare. The value of the total magnetic energy reduction ($1.17 \times 10^{32}$ erg) is the net effect of the magnetic energy increasing and decreasing in the different regions of the modeling space. Compared with the time variation of the integrated values of total magnetic energy (which is useful to evaluate the total energy budget for solar eruptive events), the characteristic of magnetic energy redistribution can provide more detailed information about the energy release process during the flare eruption.

The magnetic energy distribution is decided by the magnetic field strength distribution as shown in equation (\ref{eq:magnetic-energy}), and the strength distributions of the coronal magnetic field is related with the spatial configurations of the coronal magnetic field. In the common sense, the compact field lines correspond to strong magnetic field and high magnetic pressure, and the separating field lines correspond to the decrease of magnetic field strength and magnetic pressure. Then the magnetic energy redistribution shown in Figure 11 also demonstrates the dynamic changes of the spatial configurations of the coronal magnetic field. For example, the increase of magnetic energy in the lower space implies more compact closed field lines, and the decrease of magnetic energy in the higher space implies a relaxation of the magnetic field.

\citet{JingEA_ApJ_2008} used magnetic shear to investigate the spatial configurations of the coronal magnetic fields before and after the X3.4 flare of AR 10930. They calculated the weighted mean magnetic shear angle at each layer in a local volume above the main PIL and found that the magnetic shear after the flare increases below 8 Mm and decreases above that height. The height of 8 Mm corresponds to $11''$ ($1''\sim 714$ km), which coincides with the maximum energy density increment layer in this work (see Figure 11 and description in section 3.4.2). The relation between the variations of the magnetic energy distribution and the variations of the spatial configuration of the coronal magnetic field (represented by field lines, magnetic shear, etc.) before and after the flare deserves a detailed quantitative analysis in the future study (see also the related discussions in the papers by \citet{SchrijverEA_ApJ_2008} and \citet{JingEA_ApJ_2008}).

\setfigurenum{A1}
\begin{figure*}
  \centering
  \noindent\includegraphics[width=26pc]{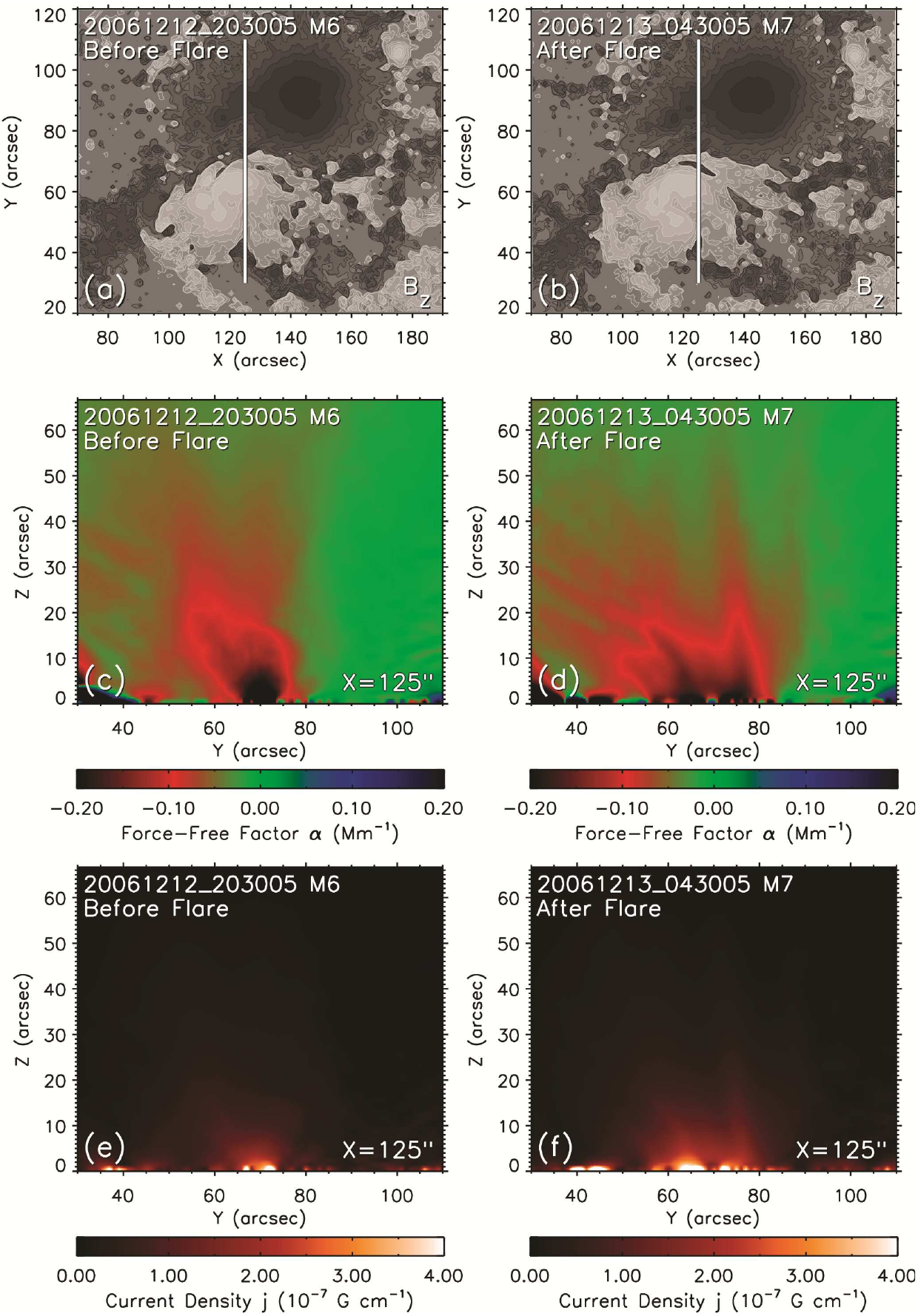}
  \caption{Diagram illustrating the vertical distributions of force-free factor $\alpha$ and current density on a slice that is parallel to the $Y$-$Z$ plane and at $125''$ on $X$ axis for the coronal magnetic field data before and after the X3.4 flare of AR 10930. (left column) Images before the flare corresponding to M6 and (right column) images after the flare corresponding to M7. The location of the slice is carefully selected to cross the breaking site of the strong magnetic connectivity described in section 3.3.3. (a and b) $B_z$ contour images of M6 and M7. The contour levels are $\pm$50, 100, 300, 500, 1000, 1500, 2000, and 3000 G. The thick white lines indicate the projected location of the selected slice on the photospheric magnetograms. (c and d) Vertical distribution images of $\alpha$ on the assigned vertical slice. The range of the color scale is set as $[-0.2, +0.2]$ Mm$^{-1}$ to display the vertical distributions of $\alpha$ more clearly, which is smaller than the color scale range adopted in other $\alpha$ figures ($[-0.3, +0.3]$ Mm$^{-1}$). (e and f) Vertical distribution images of current density on the assigned vertical slice. The absolute values of current density are employed to produce the images, which are different from the current density maps adopted in other current density figures.}
\end{figure*}

\section{Summary and Conclusion}
In this paper, the 3-D coronal magnetic field and its variations associated with the X3.4-class solar flare event of AR 10930 are studied. The time series 3-D coronal magnetic field data associated with the flare were reconstructed based on the NLFFF model and by using the extrapolation method developed in our previous work \citep{HeEA_JGR_2011, HeWang_JGR_2008}. The bottom boundaries for the NLFFF extrapolations are nine photospheric vector magnetograms obtained by the Hinode/SOT-SP instrument within the time interval of 3 days, six magnetograms were observed before the flare, and three magnetograms were observed after the flare. The reconstructed coronal magnetic field data are analyzed through three quantitative physical measures, 3-D force-free factor $\alpha$, 3-D electric current density, and 3-D magnetic energy density. We use the distribution of force-free factor $\alpha$ together with the distribution of electric current density to reflect the internal variation of the magnetic connectivity in the coronal magnetic field. For the magnetic energy analysis, we emphasize the spatial distribution variation of the 3-D magnetic energy density before and after the flare.

The distribution variations of force-free factor $\alpha$ and electric current density reveal a clear change of magnetic connectivity in the low corona of AR 10930 before and after the X3.4 flare. There exists a prominent magnetic connectivity with strong negative $\alpha$ values and strong current density magnitude before the flare which extends along the main PIL of the active region, and this strong magnetic connectivity is found to be totally broken after the flare eruption. The comparison with the flare onset imaging observation shows that the breaking site of the strong magnetic connectivity coincide with the location of the flare initial eruption, which is just over the north edge of a narrow negative flux patch that appears in the photospheric magnetogram observed just after the flare (the magnetogram labeled M7 in this paper). By examining the $\alpha$ distribution associated with M7 on the photosphere, it is found that there exists a long cramped region with positive $\alpha$ values at the north edge of the negative flux patch, which has different property of magnetic connectivity compared with the surrounding area where the negative $\alpha$ values dominate. We conclude that it is the appearance of this cramped positive $\alpha$ region that cuts off the strong magnetic connectivity before the flare and cause the broken magnetic connectivity after the flare eruption. Moreover, considering the spatial coincidence of the breaking site of the strong magnetic connectivity and the location of the flare initial eruption, we propose a scenario (sketched in Figure 15) to interpret the initial plasma ejection of the flare by the breaking of the strong magnetic connectivity. As illustrated in Figure 15, it is the Lorentz force acting on the isolated electric current at the magnetic connectivity breaking site that lifts the associated plasmas and causes the initial plasma ejection.

\setfigurenum{A2}
\begin{figure*}
  \centering
  \noindent\includegraphics[width=24pc]{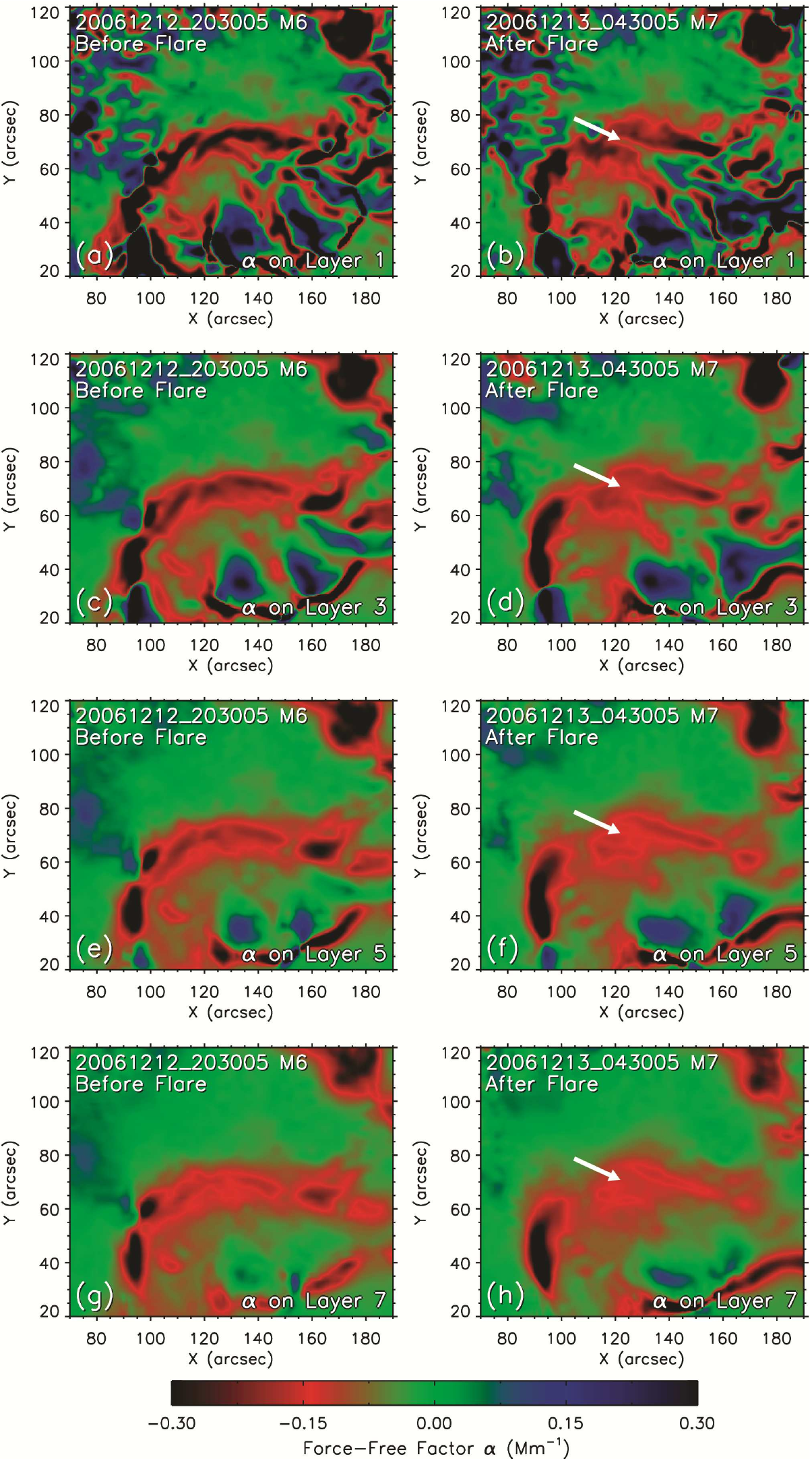}
  \caption{Distribution images of $\alpha$ on layer 1, layer 3, layer 5, and layer 7 in the central area of the active region for the coronal magnetic field data before and after the X3.4 flare of AR 10930. (left column) Images before the flare corresponding to M6 and (right column) images after the flare corresponding to M7. The white arrows in Figures A2b, A2d, A2f, and A2h are the same as the arrows shown in Figure~9, which indicate the breaking site of the strong magnetic connectivity described in section 3.3.3.}
\end{figure*}

The spatial distribution variation of the 3-D magnetic energy density in AR 10930 reveals the redistribution of magnetic energy before and after the X3.4 flare. The redistribution means the magnetic energy contained in the coronal magnetic field is not uniformly increasing or decreasing. Instead, in some regions the magnetic energy density increase and in other regions the energy density decreases. By examining the difference data of the 3-D magnetic energy density distributions just before and after the flare eruption, it is found that in the lower space of the modeling volume the increase of magnetic energy dominates and in the higher space the decrease of energy dominates. The region of magnetic energy density increasing within the lower layers is just above the main PIL of the photospheric magnetogram where the X3.4 flare happened. The comparison with the flare onset imaging observation and the distribution variations of $\alpha$ and current density shows that the site with the highest values of energy density increase at layer 11 (maximum energy increment layer) coincides with the location of the flare initial eruption as well as the breaking site of the strong magnetic connectivity mentioned above. The increase of the magnetic energy density around the breaking site of the magnetic connectivity can be interpreted by an enhancement of the magnetic field strength in the local region, which can also interpret the magnitude increase of the current density in the same region after the flare.

In the next step of this work, we plan to compare the variations of the coronal magnetic field associated with the X3.4 flare of AR 10930 with more observations obtained by various space-based and ground-based instruments \citep{MinoshimaEA_ApJ_2009, Shibasaki_ASPC_2012}, such as the UV and EUV images by the TRACE satellite \citep{HandyEA_SolPhys_1999}, soft X-ray images by the X-ray Telescope of Hinode satellite (Hinode/XRT) \citep{GolubEA_SolPhys_2007}, hard X-ray images by the Reuven Ramaty High Energy Solar Spectroscopic Imager (RHESSI) satellite \citep{LinEA_SolPhys_2002}, and the ground-based Nobeyama Radioheliograph images \citep{NakajimaEA_PIEEE_1994}, which can be helpful to having more in-depth understandings on the detailed processes of the X3.4 flare and the associated CME. The coronal physical measures employed in this work can also be the additional quantitative basis for the short-term forecasting of major solar flare event \citep{LekaBarnes_ApJ_2007, HeEA_AdSpR_2008, WangEA_RAA_2009}.



\appendix
\section{Vertical Distributions of the Force-Free Factor $\alpha$ and Electric Current Density}
A slice in the 3-D coronal data grid which is parallel to the $Y$-$Z$ plane and at $125''$ on $X$ axis is employed to illustrate the vertical distribution of force-free factor $\alpha$ and electric current density in AR 10930 before and after the X3.4 flare. The location of the slice is carefully selected to cross the breaking site of the strong magnetic connectivity described in section 3.3.3. The resulting images of the $\alpha$ and current density vertical distributions corresponding to M6 and M7 (observed just before and after the flare eruption) on the assigned slice are shown in Figure A1. To display the vertical distributions of $\alpha$ more clearly, the range of the color scale for $\alpha$ in Figures A1c and A1d is set as $[-0.2, +0.2]$ Mm$^{-1}$, which is smaller than the color scale range adopted in other $\alpha$ figures ($[-0.3, +0.3]$ Mm$^{-1}$). The vertical distribution images of current density in Figures A1e and A1f use the absolute values of current density on the assigned slice, which are different from the current density maps (see definition in section 3.3.2) adopted in other current density figures.

In the main part of this paper, we only show the horizontal distribution of $\alpha$ on layer 2. In Figure A2, we further display the $\alpha$ distributions on layer 1, layer 3, layer 5, and layer 7 of the coronal magnetic field data corresponding to M6 and M7 in the central area of the active region, which together with Figures A1c and A1d demonstrate the sharp decline of the $\alpha$ values with increase in height.

\setfigurenum{B1}
\begin{figure*}
  \centering
  \noindent\includegraphics[width=24pc]{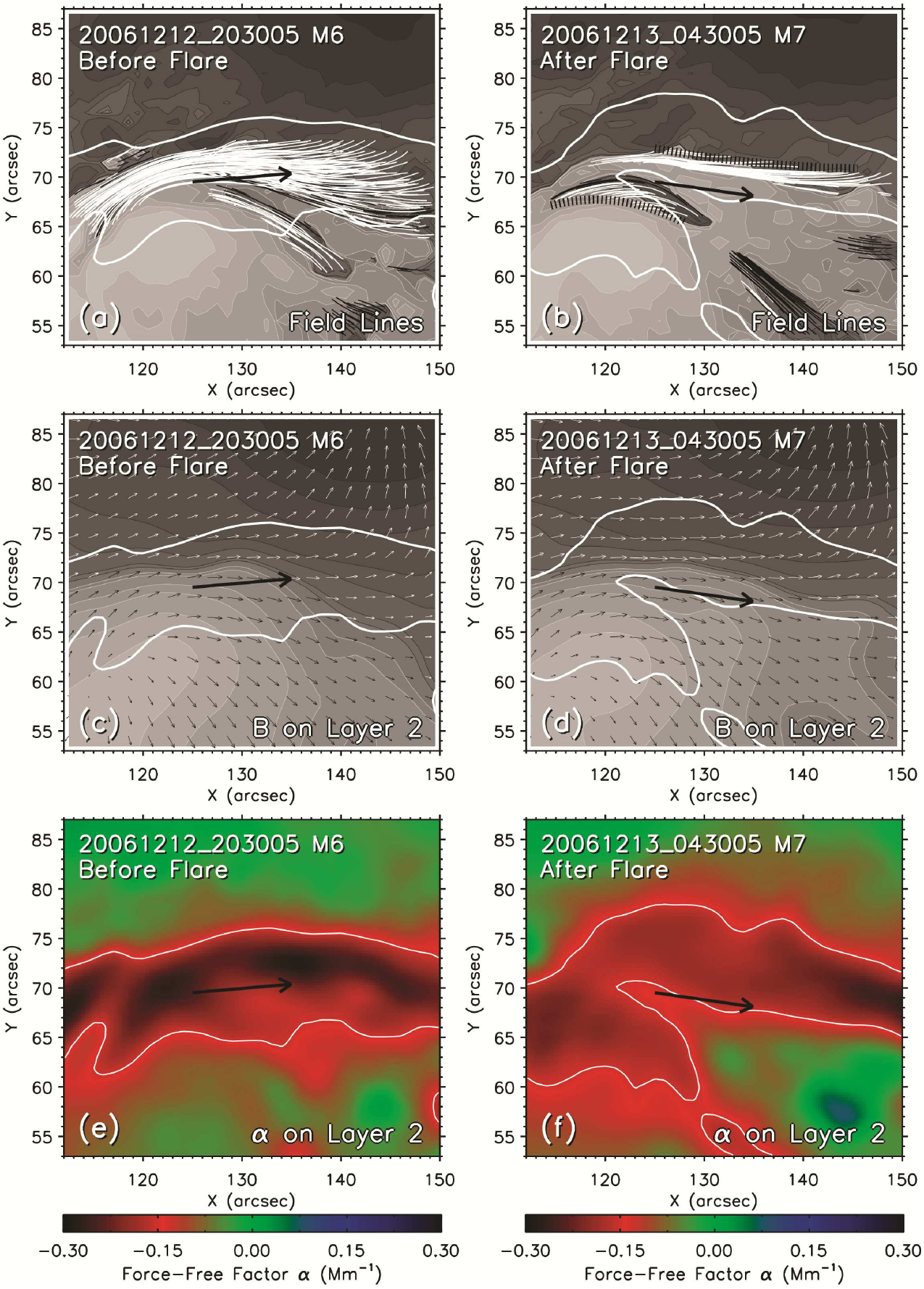}
  \caption{Magnetic field configurations associated with the strong $\alpha$ regions before and after the X3.4 flare of AR 10930. (left column) Images before the flare corresponding to M6 and (right column) images after the flare corresponding to M7. (a and b) Closed field lines images before and after the flare. The field lines are plotted overlying the $B_z$ contour images of the corresponding photospheric magnetograms. Only the closed field lines lower than layer 5 are displayed. Field lines with strong $\alpha$ values ($|\alpha|\geq0.15$ Mm$^{-1}$ or $\alpha\leq-0.15$ Mm$^{-1}$) are plotted in white color, and field lines with $|\alpha|<0.15$ Mm$^{-1}$ are plotted in black color. The two thick dotted lines in Figure B1b outline the configurations of the two field line bundles after the flare. (c and d) Vector magnetic field images on layer 2 of the reconstructed coronal magnetic field data before and after the flare. The white contours represent the positive polarity of $B_z$ on layer 2, and the black contours represent the negative polarity of $B_z$ on layer 2. The contour levels for $B_z$ are $\pm$50, 100, 300, 500, 1000, 1500, and 2000 G. Small arrows overlying the contours represent the transverse field components on layer 2. The long black arrows in Figures B1c and B1d, which correspond to M6 and M7, respectively, indicate the azimuths of the magnetic field vectors at the selected field point ($X=125'', Y=69.5''$) on layer 2. The two arrows are also plotted in other images according to their corresponding magnetograms. (e and f) Distribution images of $\alpha$ on layer 2 before and after the flare. The white contour lines in Figures B1e and B1f represent the isograms of $\alpha$ (contour level, $-0.15$ Mm$^{-1}$) that correspond to M6 and M7, respectively. The two isograms are also plotted in other images according to their corresponding magnetograms.}
\end{figure*}

\section{Magnetic Field Configurations Associated With the Strong $\alpha$ Regions Before and After the Flare}

The magnetic field configurations associated with the regions with strong $\alpha$ and strong current density values before and after the X3.4 flare of AR 10930 (corresponding to M6 and M7) are investigated through the configurations of closed field lines and the distributions of vector magnetic field on layer 2 of the reconstructed 3-D coronal magnetic field data. The resulting images in the central area of the active region and around the breaking site of the magnetic connectivity (see description in section 3.3) are shown in Figure B1. The left column in Figure B1 is for the images before the flare (corresponding to M6), and the right column is for the images after the flare (corresponding to M7). The first row in Figure B1 displays the images of the closed field lines, the second row displays the images of the vector magnetic fields on layer 2 of the reconstructed coronal magnetic field data, and the third row gives the images of the $\alpha$ distributions on layer 2 for reference.

Only the closed field lines lower than layer 5 are displayed in Figures B1a and B1b. Field lines with strong $\alpha$ values ($|\alpha|\geq0.15$ Mm$^{-1}$ or $\alpha\leq-0.15$ Mm$^{-1}$) are plotted in white color, and field lines with $|\alpha|<0.15$ Mm$^{-1}$ are plotted in black color. The isogram of the $\alpha$ distribution on layer 2 corresponding to M6 is plotted overlying the images in Figures B1a, B1c, and B1e (left column) to display the location of the strong $\alpha$ region before the flare. The isogram of $\alpha$ corresponding to M7 (after the flare) is plotted overlying the images in Figures B1b, B1d, and B1f (right column). The contour level for the $\alpha$ isograms is $-0.15$ Mm$^{-1}$.

It can be seen in Figures B1a and B1b that the closed field lines configurations can well demonstrate the distinct change of magnetic connectivity before and after the flare. In Figure B1a (before the flare) the field lines are concentrated in one bundle, in Figure B1b (after the flare) the field lines are broken into two bundles, and the orientations and locations of the field line bundles roughly coincide with the magnetic connectivity configurations manifested by the $\alpha$ distributions before and after the flare (represented by the isograms of $\alpha$ in Figure B1). In Figure B1b, we use two thick dotted lines to outline the configurations of the two bundles of field lines, which are employed in Figure 15 (see section 5.2) as the representative configuration of the broken magnetic connectivity.

In Figures B1c and B1d it can be seen that the directions of the magnetic fields in the strong $\alpha$ regions are roughly aligned along the stretching orientations of the magnetic connectivities. We select a representative field point ($X=125'', Y=69.5''$) at the breaking site of the magnetic connectivity and let the azimuths of the magnetic field vectors at this point on layer 2 as the representative directions of the magnetic connectivities in the strong $\alpha$ regions. The representative magnetic field direction for the strong magnetic connectivity before the flare is shown in Figures B1a, B1c, and B1e (left column) by long black arrows at the selected field point. The representative magnetic field direction for the broken magnetic connectivity after the flare is shown in Figures B1b, B1d, and B1f (right column) by long black arrow as well, but with southward deflection compared with the arrows in the left column of Figure B1. The representative directions for the magnetic connectivities before and after the flare (corresponding to M6 and M7) are employed in Figure 9 (see section 3.3.3) and Figure 15 (see section 5.2).

%
%

%

\begin{acknowledgments}
The authors are grateful to the reviewers for the valuable comments to improve this paper. Hinode is a Japanese mission developed and launched by ISAS/JAXA, with NAOJ as domestic partner and NASA and STFC (UK) as international partners. It is operated by these agencies in cooperation with ESA and NSC (Norway). SOHO is a project of international cooperation between ESA and NASA. Hinode/SOT-SP inversions were conducted at NCAR under the framework of the Community Spectro-Polarimetric Analysis Center (CSAC). This work is jointly supported by open fund project of Key Laboratory of Modern Astronomy and Astrophysics (Nanjing University), National Natural Science Foundation of China (NSFC) through grants 40890161 and 40890160, 11303051, 11273031, 10921303, 11025314, and 10673004, and National Basic Research Program of China (973 Program) through grant 2011CB811406. H.~He and H.~Wang also acknowledge the supports of China Meteorological Administration (grant GYHY201106011) and Strategic Priority Research Program on Space Science, Chinese Academy of Sciences (grant XDA04060801).
\end{acknowledgments}

%

%
%
\end{article}


\begin{thebibliography}{}

\providecommand{\natexlab}[1]{#1}
\expandafter\ifx\csname urlstyle\endcsname\relax
  \providecommand{\doi}[1]{doi:\discretionary{}{}{}#1}\else
  \providecommand{\doi}{doi:\discretionary{}{}{}\begingroup
  \urlstyle{rm}\Url}\fi

\bibitem[{\textit{Aschwanden}(2005)}]{Aschwanden_Book_2005}
Aschwanden, M.~J. (2005), \textit{Physics of the Solar Corona: An Introduction With Problems and Solutions}, 2nd ed., 892 pp., Praxis, Chichester, U. K.

\bibitem[{\textit{Aschwanden}(2012)}]{Aschwanden_SSRv_2012}
Aschwanden, M. J. (2012), GeV particle acceleration in solar flares and ground level enhancement (GLE) events, \textit{Space Sci. Rev.}, \textit{171}, 3--21, \doi{10.1007/s11214-011-9865-x}.

\bibitem[{\textit{Benz}(2008)}]{Benz_LRSP_2008}
Benz, A.~O. (2008), Flare observations, \textit{Living Rev. Sol. Phys.}, \textit{5}, 1, \doi{10.12942/lrsp-2008-1}.

\bibitem[{\textit{Cargill}(2009)}]{Cargill_SSRv_2009}
Cargill, P.~J. (2009), Coronal magnetism: Difficulties and prospects, \textit{Space Sci. Rev.}, \textit{144}, 413--421, \doi{10.1007/s11214-008-9446-9}.

\bibitem[{\textit{Centeno et~al.}(2009)}]{CentenoEA_ASPC_2009}
Centeno, R., B.~Lites, A.~G. de Wijn, and D.~Elmore (2009), Hinode's SP and G-band co-alignment, in \textit{The Second Hinode Science Meeting: Beyond Discovery---Toward Understanding}, \textit{Astronomical Society Of The Pacific Conference Series}, vol. 415, edited by B.~Lites et~al., pp. 323--326, Astronomical Society of the Pacific, San Francisco, Calif.

\bibitem[{\textit{Chen}(2011)}]{Chen_LRSP_2011}
Chen, P.~F. (2011), Coronal mass ejections: Models and their observational basis, \textit{Living Rev. Sol. Phys.}, \textit{8}, 1, \doi{10.12942/lrsp-2011-1}.

\bibitem[{\textit{Chen et~al.}(2011)}]{ChenEA_ApJ_2011}
Chen, Y., S.~W. Feng, B.~Li, H.~Q. Song, L.~D. Xia, X.~L. Kong, and X.~Li (2011), A coronal seismological study with streamer waves, \textit{Astrophys. J.}, \textit{728}, 147, \doi{10.1088/0004-637X/728/2/147}.

\bibitem[{\textit{DeRosa et~al.}(2009)}]{DeRosaEA_ApJ_2009}
DeRosa, M.~L., et~al. (2009), A critical assessment of nonlinear force-free field modeling of the solar corona for active region 10953, \textit{Astrophys. J.}, \textit{696}, 1780--1791, \doi{10.1088/0004-637X/696/2/1780}.

\bibitem[{\textit{Domingo et~al.}(1995)}]{DomingoEA_SolPhys_1995}
Domingo, V., B.~Fleck, and A.~I. Poland (1995), The SOHO mission: An overview, \textit{Sol. Phys.}, \textit{162}, 1--37, \doi{10.1007/BF00733425}.

\bibitem[{\textit{Fan}(2009)}]{Fan_LRSP_2009}
Fan, Y. (2009), Magnetic fields in the solar convection zone, \textit{Living Rev. Sol. Phys.}, \textit{6}, 4, \doi{10.12942/lrsp-2009-4}.

\bibitem[{\textit{Fisher and Welsch}(2008)}]{FisherWelsch_ASPC_2008}
Fisher, G.~H., and B.~T. Welsch (2008), FLCT: A fast, efficient method for performing local correlation tracking, in \textit{Subsurface and Atmospheric Influences on Solar Activity}, \textit{Astronomical Society Of the Pacific Conference Series}, vol. 383, edited by R.~Howe et~al., pp. 373--380, Astronomical Society of the Pacific, San Francisco, Calif.

\bibitem[{\textit{Fletcher et~al.}(2011)}]{FletcherEA_SSRv_2011}
Fletcher, L. et al. (2011), An observational overview of solar flares, \textit{Space Sci. Rev.}, \textit{159}, 19--106, \doi{10.1007/s11214-010-9701-8}.

\bibitem[{\textit{Fuhrmann et~al.}(2011)}]{FuhrmannEA_AA_2011}
Fuhrmann, M., N.~Seehafer, G.~Valori, and T.~Wiegelmann (2011), A comparison of preprocessing methods for solar force-free magnetic field extrapolation, \textit{Astron. Astrophys.}, \textit{526}, A70, \doi{10.1051/0004-6361/201015453}.

\bibitem[{\textit{Georgoulis}(2005)}]{Georgoulis_ApJ_2005}
Georgoulis, M.~K. (2005), A new technique for a routine azimuth disambiguation of solar vector magnetograms, \textit{Astrophys. J.}, \textit{629}, L69--L72, \doi{10.1086/444376}.

\bibitem[{\textit{Georgoulis et~al.}(2012)}]{GeorgoulisEA_ApJ_2012}
Georgoulis, M.~K., K.~Tziotziou, and N.-E. Raouafi (2012), Magnetic energy and helicity budgets in the active-region solar corona. II. Nonlinear force-free approximation, \textit{Astrophys. J.}, \textit{759}, 1, \doi{10.1088/0004-637X/759/1/1}.

\bibitem[{\textit{Golub}(2007)}]{GolubEA_SolPhys_2007}
Golub, L. et al. (2007), The X-Ray Telescope (XRT) for the Hinode mission, \textit{Sol. Phys.}, \textit{243}, 63--86, \doi{10.1007/s11207-007-0182-1}.

\bibitem[{\textit{Handy et~al.}(1999)}]{HandyEA_SolPhys_1999}
Handy, B. N. et al. (1999), The transition region and coronal explorer, \textit{Sol. Phys.}, \textit{187}, 229--260, \doi{10.1023/A:1005166902804}.

\bibitem[{\textit{He and Wang}(2006)}]{HeWang_MNRAS_2006}
He, H., and H.~Wang (2006), The validity of the boundary integral equation for magnetic field extrapolation in open space above a spherical surface, \textit{Mon. Not. R. Astron. Soc.}, \textit{369}, 207--215, \doi{10.1111/j.1365-2966.2006.10288.x}.

\bibitem[{\textit{He and Wang}(2008)}]{HeWang_JGR_2008}
He, H., and H.~Wang (2008), Nonlinear force-free coronal magnetic field extrapolation scheme based on the direct boundary integral formulation, \textit{J. Geophys. Res.}, \textit{113}, A05S90, \doi{10.1029/2007JA012441}.

\bibitem[{\textit{He et~al.}(2008)}]{HeEA_AdSpR_2008}
He, H., H.~Wang, Z.~Du, R.~Li, Y.~Cui, L.~Zhang, and Y.~He (2008), Solar activity prediction studies and services in NAOC, \textit{Adv. Space Res.}, \textit{42}, 1450--1456, \doi{10.1016/j.asr.2007.02.068}.

\bibitem[{\textit{He et~al.}(2011)}]{HeEA_JGR_2011}
He, H., H.~Wang, and Y.~Yan (2011), Nonlinear force-free field extrapolation of the coronal magnetic field using the data obtained by the Hinode satellite, \textit{J. Geophys. Res.}, \textit{116}, A01101, \doi{10.1029/2010JA015610}.

\bibitem[{\textit{Hudson}(2011)}]{Hudson_SSRv_2011}
Hudson, H.~S. (2011), Global properties of solar flares, \textit{Space Sci. Rev.}, \textit{158}, 5--41, \doi{10.1007/s11214-010-9721-4}.

\bibitem[{\textit{Inoue et~al.}(2012)}]{InoueEA_ApJ_2012}
Inoue, S., D.~Shiota, T.~T. Yamamoto, V.~S. Pandey, T.~Magara, and G.~S. Choe (2012), Buildup and release of magnetic twist during the X3.4 solar flare of 2006 December 13, \textit{Astrophys. J.}, \textit{760}, 17, \doi{10.1088/0004-637X/760/1/17}.

\bibitem[{\textit{Jiang et~al.}(2012)}]{JiangEA_ApJ_2012}
Jiang, C., X.~Feng, S.~T. Wu, and Q.~Hu (2012), Study of the three-dimensional coronal magnetic field of active region 11117 around the time of a confined flare using a data-driven CESE-MHD model, \textit{Astrophys. J.}, \textit{759}, 85, \doi{10.1088/0004-637X/759/2/85}.

\bibitem[{\textit{Jing et~al.}(2008)}]{JingEA_ApJ_2008}
Jing, J., T. Wiegelmann, Y. Suematsu, M. Kubo, and H. Wang (2008), Changes of magnetic structure in three dimensions associated with the X3.4 flare of 2006 December 13, \textit{Astrophys. J.}, \textit{676}, L81--L84, \doi{10.1086/587058}.

\bibitem[{\textit{Jing et~al.}(2010)}]{JingEA_ApJ_2010}
Jing, J., C. Tan, Y. Yuan, B. Wang, T. Wiegelmann, Y. Xu, and H. Wang (2010), Free magnetic energy and flare productivity of active regions, \textit{Astrophys. J.}, \textit{713}, 440--449, \doi{10.1088/0004-637X/713/1/440}.

\bibitem[{\textit{Kosugi et~al.}(2007)}]{KosugiEA_SolPhys_2007}
Kosugi, T., et~al. (2007), The Hinode (Solar-B) mission: An overview, \textit{Sol. Phys.}, \textit{243}, 3--17, \doi{10.1007/s11207-007-9014-6}.

\bibitem[{\textit{Kubo et~al.}(2007)}]{KuboEA_PASJ_2007}
Kubo, M., et~al. (2007), Hinode observations of a vector magnetic field change associated with a flare on 2006 December 13, \textit{Publ. Astron. Soc. Jpn.}, \textit{59}, S779--S784.

\bibitem[{\textit{Leka}(1999)}]{Leka_SolPhys_1999}
Leka, K. D. (1999), On the value of `$\alpha_{\mathrm{AR}}$' from vector magnetograph data. II. Spatial resolution, field of view, and validity, \textit{Sol. Phys.}, \textit{188}, 21--40, \doi{10.1023/A:1005130630873}.

\bibitem[{\textit{Leka and Barnes}(2007)}]{LekaBarnes_ApJ_2007}
Leka, K.~D., and G.~Barnes (2007), Photospheric magnetic field properties of flaring versus flare-quiet active regions. IV. A statistically significant sample, \textit{Astrophys. J.}, \textit{656}, 1173--1186, \doi{10.1086/510282}.

\bibitem[{\textit{Leka et~al.}}(2009)]{LekaEA_SolPhys_2009}
Leka, K. D., G. Barnes, A. D. Crouch, T. R. Metcalf, G. A. Gary, J. Jing, and Y. Liu (2009), Resolving the $180^\circ$ ambiguity in solar vector magnetic field data: Evaluating the effects of noise, spatial resolution, and method assumptions, \textit{Sol. Phys.}, 260, 83--108, \doi{10.1007/s11207-009-9440-8}.

\bibitem[{\textit{Lin et~al.}(2002)}]{LinEA_SolPhys_2002}
Lin, R. P. et al. (2002), The Reuven Ramaty High-Energy Solar Spectroscopic Imager (RHESSI), \textit{Sol. Phys.}, \textit{210}, 3--32, \doi{10.1023/A:1022428818870}.

\bibitem[{\textit{Lin et~al.}(2004)}]{LinEA_ApJ_2004}
Lin, H., J.~R. Kuhn, and R.~Coulter (2004), Coronal magnetic field measurements, \textit{Astrophys. J.}, \textit{613}, L177--L180, \doi{10.1086/425217}.

\bibitem[{\textit{Lites and Ichimoto}(2013)}]{LitesIchimoto_SolPhys_2013}
Lites, B.~W., and K.~Ichimoto (2013), The SP\_PREP data preparation package for the Hinode Spectro-Polarimeter, \textit{Sol. Phys.}, \textit{283}, 601--629, \doi{10.1007/s11207-012-0205-4}.

\bibitem[{\textit{Lites et~al.}(2007)}]{LitesEA_MmSAI_2007}
Lites, B., R.~Casini, J.~Garcia, and H.~Socas-Navarro (2007), A suite of community tools for spectro-polarimetric analysis, \textit{Mem. Soc. Astron. Ital.}, \textit{78}, 148--153.

\bibitem[{\textit{Lites et~al.}(2013)}]{LitesEA_SolPhys_2013}
Lites, B.~W. et~al. (2013), The Hinode Spectro-Polarimeter, \textit{Sol. Phys.}, \textit{283}, 579--599, \doi{10.1007/s11207-012-0206-3}.

\bibitem[{\textit{Liu et~al.}(2012)}]{LiuEA_ApSS_2012}
Liu, S., H.~Q. Zhang, and J.~T. Su (2012), Error analysis regarding the calculation of nonlinear force-free field, \textit{Astrophys. Space Sci.}, \textit{337}, 665--678, \doi{10.1007/s10509-011-0879-3}.

\bibitem[{\textit{Matsuzaki et~al.}(2007)}]{MatsuzakiEA_SolPhys_2007}
Matsuzaki, K., M.~Shimojo, T.~D. Tarbell, L.~K. Harra, and E.~E. Deluca (2007), Data archive of the Hinode mission, \textit{Sol. Phys.}, \textit{243}, 87--92, \doi{10.1007/s11207-006-0303-2}.

\bibitem[{\textit{Metcalf et~al.}(2006)}]{MetcalfEA_SolPhys_2006}
Metcalf, T. R. et al. (2006), An overview of existing algorithms for resolving the $180^\circ$ ambiguity in vector magnetic fields: Quantitative tests with synthetic data, \textit{Sol. Phys.}, 237, 267--296, \doi{10.1007/s11207-006-0170-x}.

\bibitem[{\textit{Min and Chae}(2009)}]{MinChae_SolPhys_2009}
Min, S., and J. Chae (2009), The rotating sunspot in AR 10930, \textit{Sol. Phys.}, \textit{258}, 203--217, \doi{10.1007/s11207-009-9425-7}.

\bibitem[{\textit{Minoshima et~al.}(2009)}]{MinoshimaEA_ApJ_2009}
Minoshima, T. et~al. (2009), Multiwavelength observation of electron acceleration in the 2006 December 13 flare, \textit{Astrophys. J.}, \textit{697}, 843--849, \doi{10.1088/0004-637X/697/1/843}.

\bibitem[{\textit{Nakajima et~al.}(1994)}]{NakajimaEA_PIEEE_1994}
Nakajima, H. et al. (1994), The Nobeyama radioheliograph, \textit{Proc. IEEE}, \textit{82}, 705--713, \doi{10.1109/5.284737}.

\bibitem[{\textit{Park et~al.}(2010)}]{ParkEA_ApJ_2010}
Park, S.-H., J. Chae, J. Jing, C. Tan, and H. Wang (2010), Time evolution of coronal magnetic helicity in the flaring active region NOAA 10930, \textit{Astrophys. J.}, \textit{720}, 1102--1107, \doi{10.1088/0004-637X/720/2/1102}.

\bibitem[{\textit{Pevtsov et~al.}(1995)}]{PevtsovEA_ApJ_1995}
Pevtsov, A.~A., R.~C. Canfield, and T.~R. Metcalf (1995), Latitudinal variation of helicity of photospheric magnetic fields, \textit{Astrophys. J.}, \textit{440}, L109--L112, \doi{10.1086/187773}.

\bibitem[{\textit{Priest and Forbes}(2002)}]{PriestForbes_AARv_2002}
Priest, E.~R., and T.~G. Forbes (2002), The magnetic nature of solar flares, \textit{Astron. Astrophys. Rev.}, \textit{10}, 313--377, \doi{10.1007/s001590100013}.

\bibitem[{\textit{R\'{e}gnier}(2007)}]{Regnier_MSAI_2007}
R\'{e}gnier, S. (2007), Nonlinear force-free field extrapolation: numerical methods and applications, \textit{Mem. Soc. Astron. Ital.}, \textit{78}, 126--135.

\bibitem[{\textit{R\'{e}gnier and Priest}(2007)}]{RegnierPriest_ApJ_2007}
R\'{e}gnier, S., and E.~R. Priest (2007), Free magnetic energy in solar active regions above the minimum-energy relaxed state, \textit{Astrophys. J.}, \textit{669}, L53--L56, \doi{10.1086/523269}.

\bibitem[{\textit{Sakurai}(1981)}]{Sakurai_SolPhys_1981}
Sakurai, T. (1981), Calculation of force-free magnetic field with non-constant $\alpha$, \textit{Sol. Phys.}, \textit{69}, 343--359, \doi{10.1007/BF00149999}.

\bibitem[{\textit{Sakurai}(1989)}]{Sakurai_SSRv_1989}
Sakurai, T. (1989), Computational modeling of magnetic fields in solar active regions, \textit{Space Sci. Rev.}, \textit{51}, 11--48, \doi{10.1007/BF00226267}.

\bibitem[{\textit{Scherrer et~al.}(1995)}]{ScherrerEA_SolPhys_1995}
Scherrer, P.~H., et~al. (1995), The Solar Oscillations Investigation---Michelson Doppler Imager, \textit{Sol. Phys.}, \textit{162}, 129--188, \doi{10.1007/BF00733429}.

\bibitem[{\textit{Schrijver et~al.}(2006)}]{SchrijverEA_SolPhys_2006}
Schrijver, C. J., M. L. DeRosa, T. R. Metcalf, Y. Liu, J. McTiernan, S. R\'{e}gnier, G. Valori, M. S. Wheatland, and T. Wiegelmann (2006), Nonlinear force-free modeling of coronal magnetic fields part I: A quantitative comparison of methods, \textit{Sol. Phys.}, \textit{235}, 161--190, \doi{10.1007/s11207-006-0068-7}.

\bibitem[{\textit{Schrijver et~al.}(2008)}]{SchrijverEA_ApJ_2008}
Schrijver, C. J. et al. (2008), Nonlinear force-free field modeling of a solar active region around the time of a major flare and coronal mass ejection, \textit{Astrophys. J.}, \textit{675}, 1637--1644, \doi{10.1086/527413}.

\bibitem[{\textit{Schwenn}(2006)}]{Schwenn_LRSP_2006}
Schwenn, R. (2006), Space weather: The solar perspective, \textit{Living Rev. Sol. Phys.}, \textit{3}, 2, \doi{10.12942/lrsp-2006-2}.

\bibitem[{\textit{Seehafer}}(1990)]{Seehafer_SolPhys_1990}
Seehafer, N. (1990), Electric current helicity in the solar atmosphere, \textit{Sol. Phys.}, \textit{125}, 219--232, \doi{10.1007/BF00158402}.

\bibitem[{\textit{Shibasaki}(2012)}]{Shibasaki_ASPC_2012}
Shibasaki, K. (2012), The flare on December 13, 2006 and the standard solar flare model, in \textit{The 3rd Hinode Science Meeting}, \textit{Astronomical Society Of The Pacific Conference Series}, vol. 454, edited by T.~Sekii et~al., pp. 315--320, Astronomical Society of the Pacific, San Francisco, Calif.

\bibitem[{\textit{Shibata and Magara}(2011)}]{ShibataMagara_LRSP_2011}
Shibata, K., and T.~Magara (2011), Solar flares: Magnetohydrodynamic processes, \textit{Living Rev. Sol. Phys.}, \textit{8}, 6, \doi{10.12942/lrsp-2011-6}.

\bibitem[{\textit{Tsuneta et~al.}(2008)}]{TsunetaEA_SolPhys_2008}
Tsuneta, S., et~al. (2008), The Solar Optical Telescope for the Hinode mission: An overview, \textit{Sol. Phys.}, \textit{249}, 167--196, \doi{10.1007/s11207-008-9174-z}.

\bibitem[{\textit{Wang et~al.}(2009)}]{WangEA_RAA_2009}
Wang, H.-N., Y.-M. Cui, and H.~He (2009), A logistic model for magnetic energy storage in solar active regions, \textit{Res. Astron. Astrophys.}, \textit{9}, 687--693, \doi{10.1088/1674-4527/9/6/007}.

\bibitem[{\textit{Watanabe et~al.}(2012)}]{WatanabeEA_SolPhys_2012}
Watanabe, K., S.~Masuda, and T.~Segawa (2012), Hinode flare catalogue, \textit{Sol. Phys.}, \textit{279}, 317--322, doi:10.1007/s11207-012-9983-y.

\bibitem[{\textit{Webb and Howard}(2012)}]{WebbHoward_LRSP_2012}
Webb, D.~F., and T.~A. Howard (2012), Coronal mass ejections: Observations, \textit{Living Rev. Sol. Phys.}, \textit{9}, 3, \doi{10.12942/lrsp-2012-3}.

\bibitem[{\textit{Wiegelmann}(2008)}]{Wiegelmann_JGRA_2008}
Wiegelmann, T. (2008), Nonlinear force-free modeling of the solar coronal magnetic field, \textit{J. Geophys. Res.}, \textit{113}, A03S02, \doi{10.1029/2007JA012432}.

\bibitem[{\textit{Wiegelmann and Sakurai}(2012)}]{WiegelmannSakurai_LRSP_2012}
Wiegelmann, T., and T. Sakurai (2012), Solar force-free magnetic fields, \textit{Living Rev. Sol. Phys.}, \textit{9}, 5, \doi{10.12942/lrsp-2012-5}.

\bibitem[{\textit{Yan and Li}(2006)}]{YanLi_ApJ_2006}
Yan, Y., and Z.~Li (2006), Direct boundary integral formulation for solar non-constant-$\alpha$ force-free magnetic fields, \textit{Astrophys. J.}, \textit{638}, 1162--1168, \doi{10.1086/499064}.

\bibitem[{\textit{Zhang et~al.}(2007)}]{ZhangEA_ApJ_2007}
Zhang, J., L.~Li, and Q.~Song (2007), Interaction between a fast rotating sunspot and ephemeral regions as the origin of the major solar event on 2006 December 13, \textit{Astrophys. J.}, \textit{662}, L35--L38, \doi{10.1086/519280}.

\bibitem[{\textit{Zhang and Low}(2005)}]{ZhangLow_ARAA_2005}
Zhang, M., and B.~C. Low (2005), The hydromagnetic nature of solar coronal mass ejections, \textit{Annu. Rev. Astron. Astrophys.}, \textit{43}, 103--137, \doi{10.1146/annurev.astro.43.072103.150602}.

\end{thebibliography}
\end{document}